\documentclass[journal]{IEEEtran}

\usepackage{amsmath, amssymb, graphicx, cite, bm}

\newtheorem{thm}{Theorem}
\newtheorem{lem}[thm]{Lemma}
\newtheorem{prop}[thm]{Proposition}
\newtheorem{cor}[thm]{Corollary}

\newtheorem{defi}[thm]{Definition}

\DeclareMathOperator*{\esssup}{ess\,sup}

\IEEEoverridecommandlockouts

\title{Information and Energy Transmission with Experimentally-Sampled Harvesting Functions}
\author{Daewon~Seo,~\IEEEmembership{Graduate Student Member,~IEEE} and
	Lav~R.~Varshney,~\IEEEmembership{Senior Member,~IEEE}
	\thanks{The authors are with the Department of Electrical and Computer Engineering and the Coordinated Science Laboratory, University of Illinois at Urbana-Champaign, Urbana, IL 61801, USA (e-mail: \{dseo9, varshney\}@illinois.edu).}
}

\begin{document}

\maketitle

\begin{abstract}
This paper considers the problem of simultaneous information and energy transmission (SIET), where the energy harvesting function is only known experimentally at sample points, e.g., due to nonlinearities and parameter uncertainties in harvesting circuits. We investigate the performance loss due to this partial knowledge of the harvesting function in terms of transmitted energy and information. In particular, we assume harvesting functions are a subclass of Sobolev space and consider two cases, where experimental samples are either taken noiselessly or in the presence of noise. Using constructive function approximation and regression methods for noiseless and noisy samples respectively, we show that the worst loss in energy transmission vanishes asymptotically as the number of samples increases. Similarly, the loss in information rate vanishes in the interior of the energy domain, however, does not always vanish at maximal energy. We further show the same principle applies in multicast settings such as medium access in the Wi-Fi protocol. We also consider the end-to-end source-channel communication problem under source distortion constraint and channel energy requirement, where distortion and harvesting functions both are known only at samples.
\end{abstract}

\section{Introduction}
There is growing interest in simultaneous information and energy transmission (SIET) where a single patterned energy signal carries both over a noisy channel. Information-theoretic investigation in this direction started in \cite{Varshney2008}, and has now spawned hundreds of results in the wireline \cite{Varshney2012} and especially the wireless setting (referred to as SWIPT (simultaneous wireless information and power transmission) in literature), see e.g.~\cite{ZhangMH2015, ClerckxZSNKP2019} for recent surveys. These classes of problems are important for sensor networks, Internet of Things (IoT), and similar settings where terminals may require energy.

\begin{figure}
	\centering
	\includegraphics[width=3.5in]{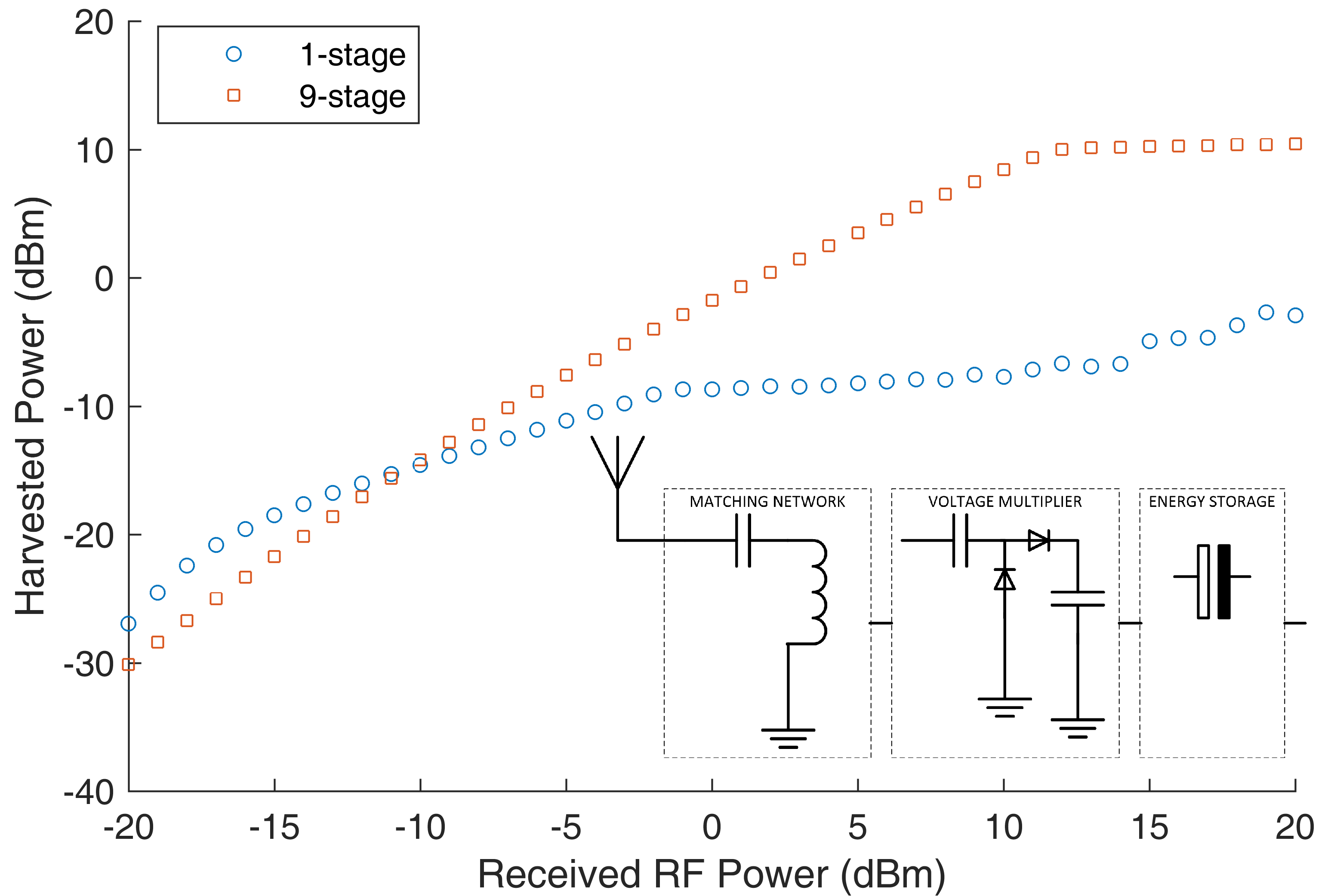}
	\caption{Samples of energy harvesting functions of the circuit shown as an inset, with one and nine stages of voltage multipliers.  Circuit design and experimental simulation data taken from \cite[Figs.~1 and 3]{NintanavongsaMLR2012}, redrawn and replotted to show relationship between RF and harvested energy.}
	\label{fig:harvesting_figure}
\end{figure}

Past theoretical works typically assume simple energy harvesting functions such as quadratic \cite{ZhouZH2013}, so the amount of energy obtained from received signal $y(t)$ is $\int_0^{\tau} y^2(t) dt$, where $\tau$ is the symbol duration. However, practical energy harvesting circuits have nonlinearities and nonidealities that complicate the relationship between channel output symbol values and their harvested energy \cite{SoyataCHsoyata2016, ValentaD2014, BoshkovskaNZS2015, KangKK2018, VarastehRC2017}. Indeed, this energy harvesting function may only be available through samples from experiments \cite{LeMF2008,StoopmanKVPS2013,StoopmanKVPS2014,SamplePSS2013,BaroudiQM2015,Zhang_nature2019} or perhaps from analog electronic circuit simulations \cite{Clerckx2018}. See Fig.~\ref{fig:harvesting_figure} for examples of harvesting circuits and their nonlinear energy harvesting functions, known only at samples \cite{NintanavongsaMLR2012}. Due to physical considerations from electromagnetics, however, we know these energy harvesting functions will be smooth in the sense of Sobolev \cite{HsiaoK1997}. Since our knowledge of harvesting functions will only be partial, it leads to a general problem of energy-requiring channel coding (and joint source-channel coding\footnote{As far as we can tell, joint source-channel coding has not been considered in the SIET literature even in the full knowledge setting.}) with partial knowledge of the energy harvesting function.

Unlike the received symbol, which is uncontrollable due to channel noise---e.g., in the low signal to noise ratio (SNR) regime, thus, it results in uncontrolled harvested energy as well---the transmitted symbol is always under control. Motivated by this limitation, unlike \cite{Varshney2008}, we think of the harvesting function as a function (or a stochastic function, e.g.~in the case of noisy measurements) of the transmitted symbol, which is a sufficiently general model for many modern communication systems.

The goal of this work is to investigate how much worst-case loss in SIET energy and information performance is incurred due to the partial knowledge of the harvesting function from samples. In particular, we study fundamental limits of point-to-point SIET systems when the signalling scheme is optimally designed based not on the full harvesting function but based on the given samples under the assumption the harvesting function is from some class of smooth functions. We consider two settings separately: when samples are noiseless or when samples are noisy. We draw on results from approximation theory including the spline method in function approximation \cite{UnserD1997} for noiseless samples, and the local polynomial estimator in non-parametric regression \cite{Tsybakov2009} for noisy samples. We prove that the worst-case amount of energy transmission is asymptotically close to the energy when the harvesting function is fully known. The worst-case information transmission is also asymptotically close in the interior of energy domain, but sampled knowledge of the harvesting function may result in full information loss in general when the system is designed for the maximum energy transmission. If the codeword is designed with a small margin away from the maximum energy transmission, it is still possible in general to achieve arbitrarily small information loss.

\begin{figure}
	\centering
	\includegraphics[width=3.5in]{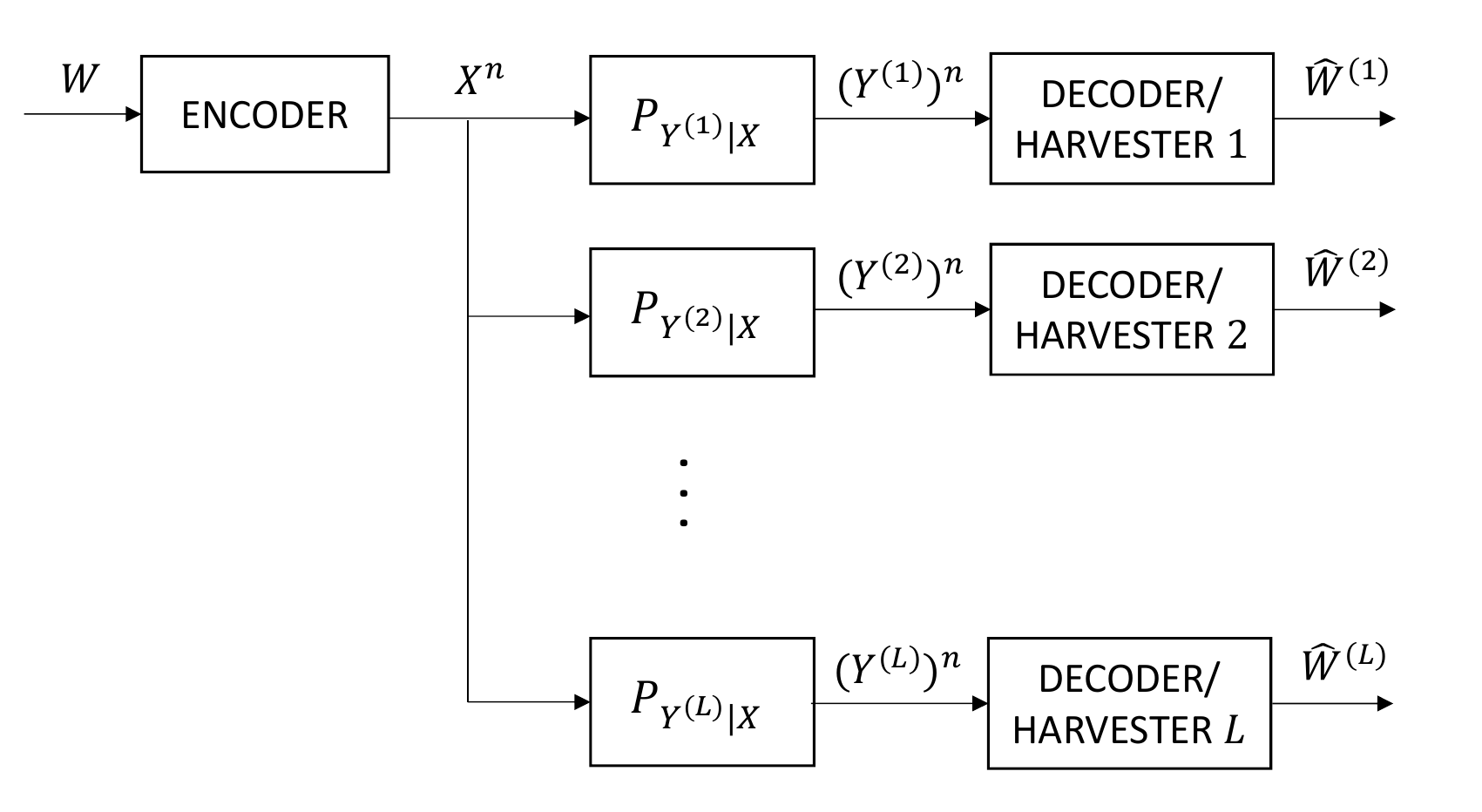}
	\caption{System model for multicast.}
	\label{fig:multicast}
\end{figure}

Moving beyond the point-to-point case, we also consider a multiterminal setting. As well as other multiterminal settings \cite{FouladgarS2012, AmorPKP2017}, a setting of medium access as in the Wi-Fi downlink protocol has been of recent interest in energy transmission using downlink Wi-Fi, but largely disconnected from optimal physical-layer designs \cite{TallaKRNSG2017}. In particular we consider multicast from a central access point, where energy and the same message are desired by several receivers, as in the beacon signal and protocol information that take up much of Wi-Fi traffic. See Fig.~\ref{fig:multicast} for a block diagram on the multicast setting, where we have different channels, harvesting functions, and energy requirements for different receiver nodes.  We find that the energy and information asymptotics from the point-to-point setting continue to hold for multicast.

Returning to the point-to-point setting, we also consider end-to-end transmission with both source and channel coding.  As far as we know, such joint source and channel coding (JSCC) problems have remained unstudied in the SIET literature, even under full information on the distortion function for lossy source coding and the energy harvesting function.  Here, we consider the problem with samples for the distortion and harvesting functions.  We build on results for lossy source coding with a sampled distortion function due to Niesen et al.~\cite{NiesenSW2006}, and make use of similar proof techniques. Since the distortion loss in source coding (\cite{NiesenSW2006}) and energy harvesting loss in SIET (Sec.~\ref{sec:SIET approx}) both asymptotically vanish, one might expect the performance loss in the end-to-end problem to also vanish asymptotically. We clarify conditions for which the loss vanishes and also give an example where the loss is bounded away from zero irrespective of the number of samples. This is important to note for end-to-end system design.

The rest of this paper is organized as follows. Sec.~\ref{sec:formulation} formally defines the unicast problem. Sec.~\ref{sec:SIET approx} studies energy and information losses incurred due to the lack of full knowledge of the true energy function for point-to-point communication. Sec.~\ref{sec:multicast} extends results to multicast. Sec.~\ref{sec:JSCC_approx} considers the end-to-end transmission problem with source distortion and channel harvesting functions. Sec.~\ref{sec:conclusion} concludes.

\section{Problem Formulation} \label{sec:formulation}
Consider the now-standard formulation of SIET systems from \cite{Varshney2008}, where the goal is to use a patterned energy signal to simultaneously transmit reliable information and energy over a noisy channel.  Recall that in a standard SIET system, first at the transmitter, messages are encoded into a codeword $x^n \in \mathcal{X}^n$ to protect against channel noise, where $n$ is codeword length. Then, the codeword is modulated into a sequence of $n$ baseband signals using a given modulation scheme, and then up-converted into a sequence of physical radio frequency (RF) waves. Attenuation and noise corrupt the RF waves so that the receiver observes a noisy version of RF waves, which is denoted by $Y^n \in \mathcal{Y}^n$. The receiver repeats the process in reverse, that is, down-converts into a baseband signal, demodulates, and decodes.

The received RF signal is also passed through an energy-harvesting circuit as in Fig.~\ref{fig:harvesting_figure}---either directly or through a signal splitting architecture \cite{Varshney2010,ZhouZH2013}---to capture energy. We suppose the information decoder and energy harvester both process the same signal. Our mathematical formulation subsumes a signal splitting scheme with a certain ratio $\rho$, called static power splitting \cite{ZhouZH2013}, with proper scaling of harvesting function. Since the receiver obtains energy from the received RF signal, in addition to maximizing information transmission between the transmitter and the receiver, a guarantee on the amount of energy delivery, say $B$, via the RF signal is also required.

\begin{figure}
	\centering
	\includegraphics[width=2.5in]{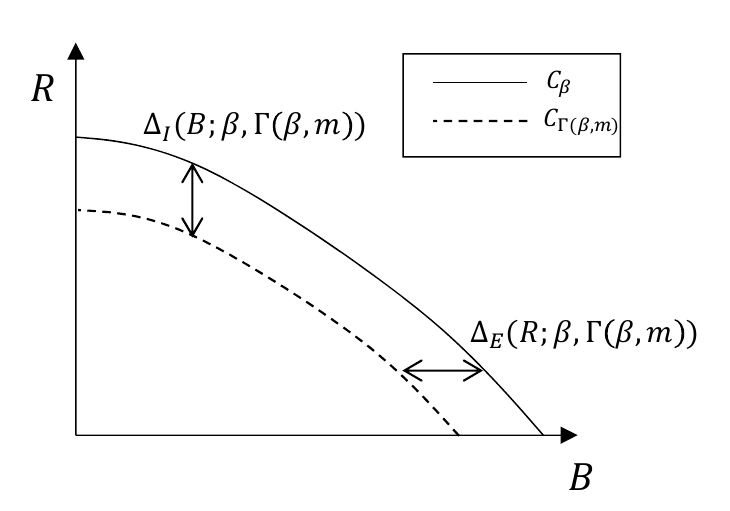}
	\caption{Typical $C_{\beta}, C_{\Gamma(\beta,m)}$ curves are depicted. Two losses $\Delta_{\textsf{E}}, \Delta_{\textsf{I}}$ incurred by sampling are defined in Sec.~\ref{subsec:loss_def}. }
	\label{fig:delta_definition}
\end{figure}

As shown in \cite{Varshney2008}, the fundamental limits of this problem are governed by the \textit{capacity-energy} function:
\begin{align}
C_b(B) = \max_{P_{X}: \mathbb{E}[b(Y)] \ge B} I(X;Y), \label{eq:Lav_eq}
\end{align}
where $X \in \mathcal{X}, Y \in \mathcal{Y}$ are transmitted and received symbols, respectively, and $b(Y)$ is the energy harvesting function for the received symbol $Y$. Note that the minimum energy requirement of \eqref{eq:Lav_eq} can be also written in terms of $x$ using conditional expectation, i.e., letting $\beta(x) := \mathbb{E}_{Y|x}[b(Y)]$,
\begin{align*}
\mathbb{E}_Y[b(Y)] = \mathbb{E}_X \left[ \mathbb{E}_{Y|X}[b(Y)] \right] = \mathbb{E}_X [\beta(X)].
\end{align*}
Hence we can think of the harvesting function as a (perhaps random) function\footnote{We assume $\beta$ is experimentally available at sample points, e.g.~by performing multiple measurments and averaging them at each point. The average corresponds to noiseless samples in Sec.~\ref{subsec:noiseless} when it is sufficiently accurate, otherwise noisy in Sec.~\ref{subsec:noisy}.} of the transmission alphabet symbols, with the following equivalent capacity-energy expression: for a harvesting function $f$ and a set of harvesting functions $F$,
\begin{align}
C_{f}(B) &= \max_{P_X: \mathbb{E}[f(X)] \ge B} I(X;Y), \label{eq:C_f} \\
C_{F}(B) &= \sup_{P_X: \mathbb{E}[f(X)] \ge B ~ \forall f \in F} I(X;Y), \label{eq:C_F}
\end{align}
which are used throughout the sequel. $C_F(B)$ indicates the maximal information rate at which we can send energy no smaller than $B$ for \emph{any} harvesting function in $F$. Note that $C_F(B) \le C_f(B)$ since the underlying probability space of \eqref{eq:C_F} is a subset to that of \eqref{eq:C_f}. As illustrated in Fig.~\ref{fig:delta_definition}, the tradeoff is non-increasing and concave.

We also define \emph{energy-capacity} functions $B_{f}(R)$, $B_{F}(R)$ as
\begin{align}
B_{f}(R) &= \max_{P_X:I(X;Y) \ge R} \mathbb{E}[f(X)], \label{eq:B_f} \\
B_{F}(R) &= \max_{P_X:I(X;Y) \ge R} \inf_{f \in F} \mathbb{E}[f(X)]. \label{eq:B_F}
\end{align}
Clearly, $B_{f}(R), B_{F}(R)$ are dual optimization problems of $C_{f}(B), C_{F}(B)$.

A probability distribution for $X$ that achieves $C_f(B)$ is called a \textit{capacity-achieving distribution}, i.e.,
\begin{align*}
P_X^* \in \arg \max_{P_X: \mathbb{E}[f(X)] \ge B} I(X;Y),
\end{align*}
where `$\in$' indicates that such capacity-achieving distribution is not necessarily unique. The maximizers with respect to $C_F(B), B_f(R), B_F(R)$ are similarily defined and also called capacity-achieving distributions. In this case, the constraint function (or set) will be clear from context. Also note that when a certain $P_X$ is given, it can be thought of as Shannon's random codebook with rate $I(X;Y)$, generated from $P_X$ \cite{CoverT1991}.

\subsection{Channel Alphabets}
In this work, we take $\mathcal{X} = [0,1]$ and $\mathcal{Y}$ as the set of all possible received signals, as determined by the physics of the system.  Taking the input alphabet as the unit interval rather than the real line imposes a peak power constraint \cite{Smith1971, Varshney2008, Varshney2012} and is motivated by practical discrete-time analog or dense constellation digital communication systems, as follows.
\begin{itemize}
\item AWGN channel: The standard AWGN channel has $\mathcal{X} = \mathcal{Y} = \mathbb{R}$ and codewords $x^n \in \mathbb{R}^n$. However, due to limitations on RF front end, we may assume $\mathcal{X} = [-a, a]$ so it is possible to assume $\mathcal{X} = [0,1]$ without loss of generality.
	
\item AM in discrete-time: In amplitude modulation (AM), at each time slot analog information $x \in [0,1] = \mathcal{X}$ is modulated and up-converted to $x \cos(2\pi f_c t)$, where $f_c$ is the carrier frequency.

\item Dense constellation QAM: Although the constellation set is discrete in $2$-dimensional space, it can be thought of as a $2$-dimensional continuous interval when sufficiently dense, say $[0,1]^2 = \mathcal{X}^2$. As an example, in dense quadratic amplitude modulation (QAM), a constellation point $\mathbf{x} =[x_1, x_2] \in [0,1]^2$ generates the RF wave $x_1 \sin(2\pi f_c t) - x_2 \cos(2\pi f_c t)$.
	
\item Dense constellation OFDM: Consider a binary sequence of length $2N$, $\mathbf{x} = [x_1, x_2, \ldots, x_{2N}] \in \{0,1\}^{2N}$. Using a $2N$-bit binary representation of real values in $[0,1]$, it can be thought of as $\{0,1\}^{2N} \approx [0,1] = \mathcal{X}$ when $N$ is large enough. Once $\mathbf{x} = [x_1, \ldots, x_{2N}] \in \mathcal{X}$ is chosen, the generated baseband signal is $\sum_{k=1}^{N} x_{2k-1} \sin(2\pi k t/T) - x_{2k} \cos(2\pi k t/T)$.
	
\item Dense constellation DSSS: Similar to OFDM, we can assume $\mathbf{x} = [x_1, x_2, \ldots, x_{2N}] \in \mathcal{X} \approx [0,1]$. Each bit of $\mathbf{x}$ is XORed with an assigned pseudo-noise (PN) sequence.
\end{itemize}

\subsection{Continuity}
We make two continuity assumptions. The first is to assume that the channel is continuous in the sense that when $x_1, x_2 \in \mathcal{X}$ are close, the distributions of $Y_1$ and $Y_2$ are also close. More precisely, when a sequence $x_n \to x$, the resulting received signals $Y_n \to Y$ in distribution.\footnote{This makes particular sense when noise is signal-independent, such as in OFDM or DSSS, where a set of length-$2N$ binary sequences in examples above can be rearranged in a Gray code manner so two successive elements differ only in one bit out of $2N$ bits. Then the one-bit difference results in RF signals that also differ only by one subcarrier element in OFDM and one PN sequence duration in DSSS, respectively. Due to the independence of noise, received signals are also similarly distributed so that the channel is continuous in the above sense.}\footnote{Note that this notion of continuity has nothing to do with capacity-achieving input distributions and their discreteness \cite{Smith1971}. Such discreteness does appear in the conditions for Thm.~\ref{thm:delta_E_LB}.} The second is to assume the energy harvesting function $\beta(\cdot)$ is smooth on $\mathcal{X}$, due to physical continuity of electromagnetic signals and circuits \cite{HsiaoK1997}. To define the smoothness rigorously, let us first introduce the $L_q$ norm and the Sobolev space $\mathcal{W}_q^{\lambda}$.
\begin{defi}
For a Lebesgue-measurable function $f$ on $\mathcal{X}$, let the $L_q$ norm for $q \in [1, \infty]$ be
\begin{align*}
|| f ||_q = \begin{cases}
\left( \int_{\mathcal{X}} |f(x)|^q dx \right)^{1/q} & \mbox{if } 1 \le q < \infty, \\
\esssup_{x \in \mathcal{X}} |f(x)| & \mbox{if } q = \infty.
\end{cases}
\end{align*}
\end{defi}
Let $\mathcal{L}_q = \mathcal{L}_q(\mathcal{X})$ be the set of all $L_q$-integrable functions on $\mathcal{X}$, i.e., $|| f ||_q < \infty$ if $f \in \mathcal{L}_q$.

\begin{defi}
For $\lambda \in \mathbb{N}, q \in [1, \infty]$, the Sobolev space $\mathcal{W}_q^{\lambda}(\mathcal{X})$ is defined as the set of functions in $\mathcal{L}_q$ such that derivatives of order equal or less than $\lambda$ exist and are in $\mathcal{L}_q$, i.e.,
\begin{align*}
\mathcal{W}_q^{\lambda}(\mathcal{X}) := \{f \in \mathcal{L}_q(\mathcal{X}) : f^{(k)} \in \mathcal{L}_q ~~~ \forall k \le \lambda \},
\end{align*}
where $f^{(k)}$ is the $k$th derivative of $f$.
\end{defi}
 
We define our class of energy harvesting functions, $\Gamma^K$, as a subset of  $\mathcal{W}_{\infty}^{\lambda}(\mathcal{X})$ satisfying:
\begin{align*}
\Gamma^K = \{\beta \in \mathcal{W}_{\infty}^{\lambda}(\mathcal{X}): ||\beta^{(k)}||_{\infty} \le K ~~~ \forall k \le \lambda \}.
\end{align*}

When the argument of $|| \cdot ||_q$ for $q \in [1, \infty]$ is a real-valued vector $\mathbf{x} \in \mathbb{R}^d$, $|| \mathbf{x} ||_q$ denotes the $\ell_q$ norm with slight abuse of notation.
\begin{align*}
|| \mathbf{x} ||_q = \begin{cases}
\left( \sum_{i=1}^d |x_i|^q \right)^{1/q} & \mbox{if } 1 \le q < \infty, \\
\max_{1 \le i \le d} |x_i| & \mbox{if } q = \infty.
\end{cases}
\end{align*}

\subsection{Sampling and Losses} \label{subsec:loss_def}

We consider \textit{regular fixed design} of samples, that is, $m$ samples are evenly-spaced on $\mathcal{X} = [0,1]$ so that $x_i = \frac{i}{m-1}$ where $i=0,1,\ldots,m-1$. Energy samples are experimentally taken either in the absence of noise or in the presence of noise, which yield different strategies. However, the choice of strategy does not make a substantial difference as we will see.

For noiseless samples $\{(\frac{i}{m-1}, \beta(\frac{i}{m-1}))\}_{i=0}^{m-1}$, let $\Gamma(\beta, m) \subset \Gamma^K$ be the set of harvesting functions that agree on the sample points. Upon observing samples, one takes a conservative strategy to transmit energy no smaller than $B$ for any harvesting function in $\Gamma(\beta, m)$. In other words, one seeks the codebook that achieves $C_{\Gamma(\beta, m)}(B)$. 

So for a given $\beta$, the energy and information losses incurred by partial knowledge are defined as
\begin{align}
\Delta_{\textsf{E}}(R; \beta, \Gamma(\beta, m)) &= B_{\beta}(R) - B_{\Gamma(\beta, m)}(R), \nonumber \\
\Delta_{\textsf{I}}(B; \beta, \Gamma(\beta, m)) &= C_{\beta}(B) - C_{\Gamma(\beta, m)}(B), \label{eq:delta_i}
\end{align}
and since the true $\beta$ is unknown, we take supremum over harvesting function in case of energy loss.

\begin{align}
\Delta_{\textsf{E}}(R) &= \sup_{\beta \in \Gamma^K} \Delta_{\textsf{E}}(R; \beta, \Gamma(\beta, m)). \label{eq:delta_e}
\end{align}
However, we do not take supremum for information loss and consider \eqref{eq:delta_i} for two reasons: energy ranges are different depending on harvesting functions, and taking supremum for information loss conceals an important insight from Thm.~\ref{thm:delta_I} and Cor.~\ref{cor:delta_I_Lip}.

For noisy samples, we assume i.i.d.\ additive measurement noise $Z_i$ with mean zero and variance $\sigma^2$ so that samples are $\{(\frac{i}{m-1}, \beta(\frac{i}{m-1}) + Z_i)\}_{i=0}^{m-1}$. Since samples are noisy, unlike noiseless samples, one cannot certify the set of true harvesting functions and design codebook for all functions in the set. Hence, one reconstructs $\hat{\beta}_m$ as accurately as possible and designs the codebook as if $\hat{\beta}_m$ is the true harvesting function. Noting that $\hat{\beta}_m$ depends on observational noise as well as $\beta$, we know that $\hat{\beta}_m$ is a stochastic mapping from $\beta$. Those facts lead us to the expected losses and minimax definition in case of energy loss as follows, where the expectations are over sample noise.
\begin{align}
&\bar{\Delta}_{\textsf{E}}(R; \beta, \hat{\beta}_m) = \mathbb{E} \left[ |B_{\beta}(R) - B_{\hat{\beta}_m}(R)| \right], \nonumber \\
&\bar{\Delta}_{\textsf{I}}(B; \beta, \hat{\beta}_m) = \mathbb{E} \left[ |C_{\beta}(B) - C_{\hat{\beta}_m}(B)| \right], \label{eq:delta_i_exp} \\
&\bar{\Delta}_{\textsf{E}}(R) = \inf_{\hat{\beta}_m} \sup_{\beta \in \Gamma^K} \bar{\Delta}_{\textsf{E}}(R; \beta, \hat{\beta}_m). \label{eq:delta_e_exp} 
\end{align}
Notice from the definition, it is immediate that $\Delta_{\textsf{I}}(B;\beta, \Gamma(\beta,m)), \Delta_{\textsf{I}}(B;\beta, \hat{\beta}_m)$ are upper-bounded by the unconstrained capacity $C_{\textsf{max}}$, i.e., for any $B$,
\begin{align}
\Delta_{\textsf{I}}(B; \beta, \Gamma(\beta, m)), \bar{\Delta}_{\textsf{I}}(B; \beta, \hat{\beta}_m) \le C_{\textsf{max}}:=\max_{P_X} I(X;Y),
\label{eq:ub}
\end{align}
which will be shown to be tight at maximum energy.

\section{Sampling Loss in Energy and Information} \label{sec:SIET approx}
This section addresses point-to-point SIET performance losses due to $m$-sample knowledge of the harvesting function. As will be seen later, the best transmitted energy based on $\hat{\beta}_m$ is arbitrary close to that based on $\beta$, so one can still design near-optimal codewords in terms of transmitted energy. Also the speed of convergence is optimal for noiseless samples under some conditions. The loss in information due to sampled knowledge vanishes at interior points of energy transmission, however, it could be arbitrary at the maximum energy transmission, say $B_{\textsf{max}}$ for noiseless samples. Thus, a system designer needs to be careful when targeting $B_{\textsf{max}}$ or should design with a small margin away from $B_{\textsf{max}}$. We constructively propose kernel-based reconstruction for noiseless and noisy samples, yielding near-optimal performance guarantees on transmitted energy.

\subsection{Noiseless Samples} \label{subsec:noiseless}
Consider noiseless samples. Reconstructing a continuous signal from samples has been a popular topic in signal processing \cite{Unser2000, VetterliKG2014}, approximation theory \cite{DeVoreL1993}, and many other engineering fields. Among numerous reconstruction methods, consider the spline method (our converse argument in Thm.~\ref{thm:delta_E_LB} will show this to be a good choice), which has piecewise polynomials as interpolant kernels to achieve efficient implementation. Since it is a local technique, rather than a global polynomial approximation method such as Lagrange interpolation, the value of the reconstructed function $\hat{f}_m(x)$ only depends on a few neighboring samples of $x$ and numerical instability called Runge's phenomenon does not appear \cite{VetterliKG2014}. See surveys \cite{Unser2000,Unser1999} for introductory material and \cite{Boor1978} for details.

Before giving our main theorems and proofs, first recall the following result on spline reconstruction in Sobolev spaces.
\begin{lem}[Prop.~3.1 in \cite{UnserD1997}] \label{lem:spline}
For $f \in \mathcal{W}_{\infty}^{\lambda}$, let $\hat{f}_m^{\textsf{SP}} \in \Gamma(f, m)$ be the spline reconstructed function. Then, for some constant $c$,
\begin{align*}
|| f - \hat{f}_m^{\textsf{SP}} ||_{\infty} \le c m^{-\lambda} ||f^{(\lambda)}||_{\infty} ~~~ \forall f \in W_{\infty}^{\lambda}.
\end{align*}
\end{lem}

Now we give a main result, which shows one can attain near-optimal transmitted energy despite the sampled harvesting function.
\begin{thm} \label{thm:delta_E_UB}
$\Delta_{\textsf{E}}(R) = O(m^{-\lambda}) ~~~ \forall R \ge 0$.
\end{thm}
\begin{IEEEproof}
Note that the best codebooks for $B_{\beta}(R)$ and $B_{\Gamma(\beta,m)}(R)$ are not necessarily identical. However, as will be seen, any codebook performs almost the same under $\beta$ and $ \hat{\beta}_m \in \Gamma(\beta, m)$.
	
First consider an arbitrary distribution $P_X$ and Shannon's random codebook generated from it. Then,
\begin{align}
&~ \left| \mathbb{E}_{P_X} \left[ \beta(X) \right] - \mathbb{E}_{P_X} [ \hat{\beta}_m(X) ] \right| \nonumber \\
\le&~ \mathbb{E}_{P_X} \left[ | \beta(X) - \hat{\beta}_m(X) | \right] = \int_{\mathcal{X}} P_X(x) |\beta(x) - \hat{\beta}_m(x) |dx \nonumber \\
\le&~ \int_{\mathcal{X}} P_X(x) || \beta - \hat{\beta}_m ||_{\infty} dx = ||\beta - \hat{\beta}_m||_{\infty}, \label{eq:thm4_bound1}
\end{align}
where the last inequality follows from the sup-norm definition, $||\beta - \hat{\beta}_m ||_{\infty} = \esssup_{x \in \mathcal{X}} |\beta(x) - \hat{\beta}_m(x)|$. Furthermore, using the triangle inequality, we have
\begin{align*}
||\beta - \hat{\beta}_m||_{\infty} \le ||\beta - \hat{\beta}_m^{\textsf{SP}}||_{\infty} + || \hat{\beta}_m^{\textsf{SP}} - \hat{\beta}_m ||_{\infty}.
\end{align*}
The first term is bounded by $c m^{-\lambda} ||\beta^{(\lambda)}||_{\infty}$ by Lem.~\ref{lem:spline}. Furthermore, note that $\hat{\beta}_m^{\textsf{SP}}$ can be seen as a spline reconstruction for another $\beta' \in \Gamma(\beta, m)$ since $\beta, \beta'$ both agree on sample points. This means the second term is also bounded by $c m^{-\lambda} ||\beta^{(\lambda)}||_{\infty}$. Therefore, from the definition of $\Gamma^K$,
\begin{align}
\left| \mathbb{E}_{P_X} \left[ \beta(X) \right] - \mathbb{E}_{P_X} [ \hat{\beta}_m(X) ] \right| \le 2c K m^{-\lambda}. \label{eq:thm4_bound}
\end{align}
It should be noted that \eqref{eq:thm4_bound} is independent of $P_X, \beta, \hat{\beta}_m$.

Next, fix $R \ge 0$ and consider $A := \{P_X: I(X;Y) \ge R\}$. Also define two capacity-achieving distributions $P_X^*, Q_X^* \in A$ for $B_{\beta}(R), B_{\Gamma(\beta,m)}(R)$, respectively. Then, we have a chain of inequalities
\begin{align*}
B_{\beta}(R) &\stackrel{(a)}{\ge} B_{\Gamma(\beta,m)}(R) = \min_{\hat{\beta}_m \in \Gamma(\beta,m)} \mathbb{E}_{Q_X^*}[\hat{\beta}_m(X)] \\
&\stackrel{(b)}{\ge} \min_{\hat{\beta}_m \in \Gamma(\beta,m)} \mathbb{E}_{P_X^*}[\hat{\beta}_m(X)] \\
&\stackrel{(c)}{\ge} \mathbb{E}_{P_X^*}[\beta(X)] - 2c K m^{-\lambda} \\
&= B_{\beta}(R) - 2c K m^{-\lambda},
\end{align*}
where (a) follows from the definitions \eqref{eq:B_f} and \eqref{eq:B_F}, (b) follows since $P_X^*$ is suboptimal for $B_{\Gamma(\beta,m)}(R)$, and (c) follows since \eqref{eq:thm4_bound} holds for all $\beta \in \Gamma^K$ and $\hat{\beta}_m \in \Gamma(\beta, m)$. Hence, we conclude that $\Delta_{\textsf{E}}(R; \beta, \Gamma(\beta, m)) = O(m^{-\lambda})$ for all $\beta \in \Gamma^K$. Since $R$ is arbitrary and the bound does not depend on $\beta$, $\Delta_{\textsf{E}}(R) = O(m^{-\lambda})$ for all $R$.
\end{IEEEproof}

From the result, we know that the conservative transmission scheme performs near-optimally in terms of energy. However, the scheme needs optimization with respect to uncountably many $\hat{\beta}_m \in \Gamma(\beta, m)$, which does not reveal a clear  codebook design. The following corollary suggests that $\hat{\beta}_m^{\textsf{SP}}$ is a good proxy for unknown $\beta$ enabling us to design near-optimal codewords as if $\hat{\beta}_m^{\textsf{SP}}$ is the true harvesting function.
\begin{cor} \label{cor:sp_based_codebook}
Codewords designed based on $\hat{\beta}_m^{\textsf{SP}}$ achieves $O(m^{-\lambda})$ loss of transmitted energy with respect to $B_{\beta}(R)$.
\end{cor}
\begin{IEEEproof}
Fix an arbitrary $R \ge 0$ and consider $B_{\hat{\beta}_m^{\textsf{SP}}}(R), B_{\beta}(R)$. Two optimal codebooks are generated from the capacity-achieving distributions for $B_{\hat{\beta}_m^{\textsf{SP}}}(R), B_{\beta}(R)$, say $P_X^*, Q_X^*$.
	
Then, under $\beta$ the optimal codebook for $\hat{\beta}_m^{\textsf{SP}}$ (i.e., $P_X^*$) performs as:
\begin{align*}
&~ |B_{\hat{\beta}_m^{\textsf{SP}}}(R) - \mathbb{E}_{P_X^*} [ \beta(X) ] | \\
=&~ \left| \mathbb{E}_{P_X^*} \left[ \hat{\beta}_m^{\textsf{SP}}(X) \right] - \mathbb{E}_{P_X^*} [ \beta(X) ] \right| \\
\stackrel{(a)}{\le}&~ || \beta - \hat{\beta}_m^{\textsf{SP}}||_{\infty} \le c K m^{-\lambda},
\end{align*}
where (a) follows from \eqref{eq:thm4_bound}. As $P_X^*$ is suboptimal for $\beta$, we know that
\begin{align*}
B_{\beta}(R) \ge B_{\hat{\beta}_m^{\textsf{SP}}}(R) - cKm^{-\lambda}.
\end{align*}

Similarly, exchanging roles of $\beta, \hat{\beta}_m^{\textsf{SP}}$ and considering the optimal codebook for $\beta$ (i.e., $Q_X^*$) gives
\begin{align*}
|B_{\beta}(R) - \mathbb{E}_{Q_X^*} [ \hat{\beta}_m^{\textsf{SP}}(X) ] | \le || \beta - \hat{\beta}_m^{\textsf{SP}}||_{\infty} \le c K m^{-\lambda}.
\end{align*}
As $Q_X^*$ is suboptimal for $\hat{\beta}_m^{\textsf{SP}}$, we know that
\begin{align*}
B_{\hat{\beta}_m^{\textsf{SP}}}(R) \ge B_{\beta}(R) - cKm^{-\lambda}.
\end{align*}
Combining the two, we have
\begin{align*}
B_{\hat{\beta}_m^{\textsf{SP}}}(R) + cKm^{-\lambda} \le B_{\beta}(R) \le B_{\hat{\beta}_m^{\textsf{SP}}}(R) - cKm^{-\lambda}.
\end{align*}
Hence, we conclude that the codebook designed based on $\hat{\beta}_m^{\textsf{SP}}$ is nearly optimal within $O(m^{-\lambda})$.
\end{IEEEproof}

It should be noted that Thm.~\ref{thm:delta_E_UB} is not tight in general, e.g., consider a peak-power constrained AWGN channel \cite{Smith1971} and suppose the capacity-achieving distribution, which is discrete, is supported on (a part of) sample points. As $\beta, \hat{\beta}_m$ always agree on sample points, $\Delta_{\textsf{E}}(R)$ is zero. However, there are cases such that the bound in Thm.~\ref{thm:delta_E_UB} is tight. Before proceeding to demonstration, we define function-wise loss.
\begin{align*}
\Delta_{\textsf{E}}'(R; \beta, \hat{\beta}_m) &= |B_{\beta}(R) - B_{\hat{\beta}_m}(R)|, \\
\Delta_{\textsf{E}}'(R) &= \sup_{\substack{\beta \in \Gamma^K, \\ \hat{\beta}_m \in \Gamma(\beta, m)}} \Delta_{\textsf{E}}'(R; \beta, \hat{\beta}_m).
\end{align*}
\begin{lem}
	$\Delta_{\textsf{E}}'(R) \le \Delta_{\textsf{E}}(R)$.
\end{lem}
\begin{IEEEproof}
Consider the left side
\begin{align*}
\Delta_{\textsf{E}}'(R) = \sup_{\beta, \hat{\beta}_m} |B_{\beta}(R) - B_{\hat{\beta}_m}(R)|
\end{align*}
and note that $\hat{\beta}_m$ is a candidate for $\beta$, but, $\beta$ is also a candidate for $\hat{\beta}_m$ since they both agree on the sample points. Hence, we can exchange $\beta, \hat{\beta}_m$ and without loss of generality, it is sufficient to consider pairs $(\beta, \hat{\beta}_m)$ such that $B_{\beta}(R) \ge B_{\hat{\beta}_m}(R)$. For any such $(\beta, \hat{\beta}_m)$, 
\begin{align*}
B_{\beta}(R) - B_{\hat{\beta}_m}(R) \le B_{\beta}(R) - B_{\Gamma(\beta, m)}(R)
\end{align*}
by definition of $B_{\Gamma(\beta, m)}(R)$. Taking supremum over all such $(\beta, \hat{\beta}_m)$ does not change the inequality, which completes the proof.
\end{IEEEproof}

Therefore, to show the lower bound on $\Delta_{\textsf{E}}(R)$, it is sufficient to show a lower bound for $\Delta_{\textsf{E}}'(R)$. The following theorem states conditions for which $\Delta_{\textsf{E}}'(R) = \Omega(m^{-\lambda})$, i.e., the bound is tight.

\begin{thm} \label{thm:delta_E_LB}
Fix some $B \in (0, B_{\textsf{max}})$. Suppose the capacity-achieving distribution $P_X^*$ yielding the Shannon's random codebook of rate $R = C_{\hat{\beta}_m}(B)$ satisfies one of the following conditions:
\begin{enumerate}
	\item $P_X^*$ is continuous and non-vanishing on $\mathcal{X}$, i.e., $P_X^*(x)\ge c$ for some $c$.
	
	\item $P_X^*$ is supported on a finite set of mass points\footnote{The discrete distribution is particularly important because the optimal input distribution is discrete in many cases especially when $\mathcal{X}$ is compact and convex and channel noise is additive, see \cite{Smith1971, Tchamkerten2004, ElMoslimanyD2018, Varshney2012}. Also refer to  \cite{DytsoGPS2018} for general channels.} disjoint from the sample points, as specified in the proof.
\end{enumerate}
Then, $\Delta_{\textsf{E}}'(R) = \Omega(m^{-\lambda})$ at $R$.
\end{thm}
\begin{IEEEproof}
We consider $\Delta'(R;\beta, \hat{\beta}_m)$ and the lower bound can be shown by a bumpy function. Thm.~4.3 in \cite{Kudryavtsev1995} states that there exists a non-negative function $f$ such that $f(x_i) = 0$ at every $x_i$ and $||f||_1 \ge c' m^{-\lambda}$. First consider the case 1). Take $\beta, \hat{\beta}_m$ as
\begin{align*}
\beta(x) &= M ~~~ \forall x \in \mathcal{X}, \\
\hat{\beta}_m(x) &= M(1-f(x)) ~~~ \forall x \in \mathcal{X},
\end{align*}
where $M$ is a constant. Then, $B_{\beta}(R) = M$ for any codebook. Also,
\begin{align*}
&~ B_{\hat{\beta}_m}(R) = \mathbb{E} \left[ \hat{\beta}_m(X) \right] = \int_{\mathcal{X}} P_X^*(x) \hat{\beta}_m(x) dx \\
=&~ \int_{\mathcal{X}} P_X^*(x) M(1-f(x)) dx = M - M\int_{\mathcal{X}} P_X^*(x)f(x) dx \\
\le&~ M - M \int_{\mathcal{X}} c f(x) dx = M - c M ||f||_1 \\
\le&~ M - c c' M m^{-\lambda}.
\end{align*}
Thus, $\Delta_{\textsf{E}}(R;\beta, \hat{\beta}_m) = |B_{\beta}(R) - B_{\hat{\beta}_m}(R)| \ge c c' M m^{-\lambda}$. We have the desired lower bound of $\Delta_{\textsf{E}}(R)$ as $\Omega(m^{-\lambda})$.

For the case 2), we repeat the above argument with $\beta(x) = M, \hat{\beta}_m = M(1-f(x))$. Since $P_X^*$ is supported on a discrete set, say $\{x_k\}$,
\begin{align*}
B_{\beta}(R) - B_{\hat{\beta}_m}(R) &= M \int_{\mathcal{X}} P_X^*(dx) f(x) \\
&= M \sum_{k} P_X^*(x_k) f(x_k).
\end{align*}
Note that by the norm monotonicity with respect to a bounded measure, $||f||_{\infty} \ge ||f||_1 \ge c'm^{-\lambda}$,  there is a disjoint point from samples such that $f(x) \ge c'm^{-\lambda}$. So when $\{x_k\}$ satisfy $f(x_k) \ge c'm^{-\lambda}$,
\begin{align*}
B_{\beta}(R) - B_{\hat{\beta}_m}(R) \ge M \sum_{k} P_X^*(x_k) c'm^{-\lambda} = Mc'm^{-\lambda},
\end{align*}
which proves $\Delta_{\textsf{E}}(R) = \Omega(m^{-\lambda})$.
\end{IEEEproof}

The next theorem and corollary deal with the information loss incurred by sampling. As will be seen below, the loss is negligible on most of the targeted energy range, however, the trivial unconstrained capacity upper bound on $\Delta_{\textsf{I}}(B;\beta, \Gamma(\beta,m))$ given as \eqref{eq:ub} could be indeed tight at $B_{\textsf{max}}$.

\begin{thm} \label{thm:delta_I}
For any $\beta \in \Gamma^K$ and $B \in [0, B_{\textsf{max}})$,
\begin{align*}
\Delta_{\textsf{I}}(B;\beta, \Gamma(\beta,m)) \to 0 ~~ \textrm{as } m \to \infty.
\end{align*}
Furthermore, there is a pair of harvesting function and channel for which $\Delta_{\textsf{I}}(B_{\textsf{max}};\beta, \Gamma(\beta,m)) = C_{\textsf{max}}$.
\end{thm}
\begin{IEEEproof}
Let us prove the first claim. At $B=0$, note that it is the same as the unconstrained capacity, i.e., $C_{\beta}(0) = C_{\hat{\beta}_m}(0) = C_{\textsf{max}}$. So $\Delta_{\textsf{I}}(0;\beta, \Gamma(\beta,m)) = 0$.

For $B \in (0, B_{\textsf{max}})$, recall that since $C_{\beta}(B)$ is concave, it is continuous over the interior of its domain, i.e., continuous on $(0, B_{\textsf{max}})$. Thm.~\ref{thm:delta_E_UB} guarantees that for every $B$, there exists a $B'$ that attains $C_{\Gamma(\beta,m)}(B) = C_{\beta}(B')$ for some close $B, B'$ with $|B-B'| = O(m^{-\lambda})$, so that at $B \in (0, B_{\textsf{max}})$,
\begin{align*}
\Delta_{\textsf{I}}(B; \beta, \Gamma(\beta, m)) &= C_{\beta}(B) - C_{\Gamma(\beta, m)}(B) \\
&= C_{\beta}(B) - C_{\beta}(B') \\
&= C_{\beta}(B) - C_{\beta}(B+O(m^{-\lambda})).
\end{align*}
Due to the continuity of $C_{\beta}$, $\Delta_{\textsf{I}}(B; \beta, \Gamma(\beta, m)) \to 0$ as $B + O(m^{-\lambda}) \to B$. The first claim is proved.

To show the second claim, fix a large $m$. We will prove by a counterexample. Take a constant $\beta$, that is, $\beta(x) = M$ over all $x$. Then, as any $P_X$ is admissible for $B \le M$ and none is for $B > M$,
\begin{align*}
C_{\beta}(B) = \begin{cases} 
C_{\textsf{max}} & \mbox{if } B \le M \\
0 & \mbox{if } B > M.
\end{cases}
\end{align*}
However, $\Gamma(\beta, m)$ definitely has an element such that $\hat{\beta}_m(x) < \beta(x) = M$ except for given sample points. In other words, $\hat{\beta}_m < \beta$ almost everywhere, so that $\mathbb{E} [\hat{\beta}_m(X)] < M$ unless $P_X$ only has point masses on the sample points. Therefore, discrete $P_X$s are the only admissible probability distributions for the energy requirement $M(=B_{\textsf{max}})$.

For such a discrete $P_X$, consider an adversarial channel 
\begin{align*}
Y = (X+Z) \mod 1,
\end{align*}
where $Z$ is an input-dependent additive noise on $\mathcal{X} = [0,1]$. The dependency is as follows: $Z$ is uniform over $[0,1]$ when $X \in \{\frac{i}{m-1}\}_{i=0}^{m-1}$, and the probability density of $Z$ is more concentrated around $0$ as $X$ is more distant from $\{\frac{i}{m-1}\}_{i=0}^{m-1}$. Since the discrete $P_X$ only sees uniform noise, $I(X;Y)$ is zero, i.e., $C_{\Gamma(\beta, m)}(M) = 0$, however, we can send information using a non-discrete $P_X$ because noise is biased toward $0$ except for sample points. Hence, $\Delta_{\textsf{I}}(B_{\textsf{max}}; \beta, \Gamma(\beta,m)) = C_{\textsf{max}}$ for this harvesting function and channel.
\end{IEEEproof}
Since we can construct the above counterexample at any particular $B$, $\sup_{\beta \in \Gamma^K} \Delta_{\textsf{I}}(B; \beta, \Gamma(\beta,m)) = C_{\textsf{max}}$. This does not give any insight into design from samples.

Although Thm.~\ref{thm:delta_I} describes the convergence of $\Delta_{\textsf{I}}$, it does not characterize $\Delta_{\textsf{I}}$ in terms of the number of samples. As the next corollary shows, the Lipschitz continuity enables us to characterize $\Delta_{\textsf{I}}(B)$ in terms of $m$ for all $B \in (0, B_{\textsf{max}})$.
\begin{cor} \label{cor:delta_I_Lip}
Suppose the channel yields Lipschitz continuous $C_{\beta}(B)$ with Lipschitz coefficient $M$ for $\beta \in \Gamma^K$ except for its end points, i.e., for $B_1, B_2 \in (0,B_{\textsf{max}})$,
\begin{align}
|C_{\beta}(B_1) - C_{\beta}(B_2)| \le M |B_1 - B_2|. \label{eq:Lipsch_def}
\end{align}
Then, $\Delta_{\textsf{I}}(B; \beta, \Gamma(\beta,m)) = O(m^{-\lambda})$ for any $B \in [0, B_{\textsf{max}})$. 
\end{cor}
\begin{IEEEproof}
When $B=0$, it is unconstrained capacity, so $\Delta_{\textsf{I}}(0;\beta, \Gamma(\beta,m)) = 0$.

For $B \in (0, B_{\textsf{max}})$ and a given $\beta \in \Gamma^K$,
\begin{align*}
\Delta_{\textsf{I}}(B; \beta, \Gamma(\beta, m)) &= C_{\beta}(B) - C_{\Gamma(\beta, m)}(B) \\
&\le C_{\beta}(B) - C_{\beta}(B + O(m^{-\lambda})) \\
&\le M O(m^{-\lambda}) = O(m^{-\lambda}),
\end{align*}
where the last inequality follows from \eqref{eq:Lipsch_def}.
\end{IEEEproof}

Thm.~\ref{thm:delta_E_UB} and Cor.~\ref{cor:sp_based_codebook} ensure Shannon's random codebook designed for $\hat{\beta}_m^{\textsf{SP}}$ is nearly close to the optimal codebook for $\beta$ in terms of transmitted energy. Further, Thm.~\ref{thm:delta_E_LB} shows that its performance is in fact asymptotically tight under some conditions on $P_X^*$.

From the same argument, Thm.~\ref{thm:delta_I} and Cor.~\ref{cor:delta_I_Lip} both basically ensure that the codebook designed as if $\hat{\beta}_m^{\textsf{SP}}$ is the true harvesting function also delivers nearly maximal information. However, please be careful when interpreting the second statement of Thm.~\ref{thm:delta_I}. The statement does not imply the codebook fails to be decoded correctly at $B_{\textsf{max}}$; rather it means that partial knowledge of the harvesting function may lower (or set higher) the targeted information rate by a non-vanishing amount in the codebook design stage. However such a mismatched codebook is always decodable since the channel remains the same regardless of sampling. This pitfall leads a system designer to stepping back from $B_{\textsf{max}}$, i.e., setting a safety energy margin from $B_{\textsf{max}}$.

\subsection{Noisy Samples} \label{subsec:noisy}
Consider noisy samples. In particular, received signal varies even for the same transmission signal. Or the noise could be due to errors in measuring battery status. In particular, we consider i.i.d.\ additive noise $Z_i$ with mean zero and variance $\sigma^2$ so that samples are $\{(x_i, T_i)\}_{i=0}^{m-1}$, where $x_i = \frac{i}{m-1}, T_i = \beta(\frac{i}{m-1}) + Z_i$.

As a constructive reconstruction method, we consider local polynomial estimation of order $\lambda$ \cite{Tsybakov2009}, denoted by $\hat{\beta}_m^{\textsf{LP}}$, since $\Gamma^K$ is differentiable upto order $\lambda$. Consider a symmetric kernel $\phi(x)$ on $[-1,1]$ such that $|\phi(x)| \le \phi_{\textsf{max}} < \infty$ and let $h$ be bandwidth. Then, $\hat{\beta}_m^{\textsf{LP}}(x)$ for a particular $x$ is obtained from $\{w_t\}_{t=0}^{\lambda}$ that solves
\begin{align}
\min_{w_i} \sum_{i=0}^{m-1} \phi \left( \frac{x_i-x}{h} \right) \left( T_i - \sum_{t=0}^{\lambda} w_t (x_i-x)^t \right)^2. \label{eq:LP_min}
\end{align}
To express $\hat{\beta}_m^{\textsf{LP}}(x)$ in closed form, it is convenient to introduce vector and matrix representations:
\begin{align*}
\mathbf{X}_x &= \begin{bmatrix}
1 & (x_0 - x) & \cdots & (x_0-x)^{\lambda} \\
1 & (x_1 - x) & \cdots & (x_1-x)^{\lambda} \\
\vdots & \vdots & \ddots & \vdots \\
1 & (x_{m-1} - x) & \cdots & (x_{m-1}-x)^{\lambda} \\
\end{bmatrix}, \\
\mathbf{T} &= [T_0, T_1, \ldots, T_{m-1}]^T, \\
\mathbf{w} &= [w_0, w_1, \ldots, w_{\lambda}]^T, \\
\mathbf{\Phi}_x &= \begin{bmatrix}
\phi(\tfrac{x_0-x}{h}) & 0 & \cdots & 0 \\
0 & \phi(\tfrac{x_1-x}{h}) & \cdots & 0 \\
\vdots & \vdots & \ddots & \vdots \\
0 & 0 & \cdots & \phi(\tfrac{x_{m-1}-x}{h})
\end{bmatrix}.
\end{align*}
Then, \eqref{eq:LP_min} is rewritten as a least squares problem
\begin{align*}
\min_{\mathbf{w}} ( \mathbf{T} - \mathbf{X}_x \mathbf{w})^T \mathbf{\Phi}_x (\mathbf{T} - \mathbf{X}_x \mathbf{w}),
\end{align*}
and the solution to this is
\begin{align*}
\mathbf{w}^* = [w_0^*, w_1^*, \ldots, w_{\lambda}^*]^T = (\mathbf{X}_x^T \mathbf{\Phi}_x \mathbf{X}_x)^{-1} (\mathbf{X}_x^T \mathbf{\Phi}_x \mathbf{T}).
\end{align*}
Then, $\hat{\beta}_m^{\textsf{LP}}(x) = w_0$, in other words,
\begin{align}
\hat{\beta}_m^{\textsf{LP}}(x) = \mathbf{e}_1^T(\mathbf{X}_x^T \mathbf{\Phi}_x \mathbf{X}_x)^{-1} (\mathbf{X}_x^T \mathbf{\Phi}_x \mathbf{T}), \label{eq:local_poly}
\end{align}
where length-$(\lambda+1)$ vector $\mathbf{e}_1$ has a $1$ in the first coordinate and $0$s otherwise. In particular when the order is zero, it is called the Nadaraya-Watson estimator \cite{Tsybakov2009}.

\begin{lem}[Thm.~1.6 in \cite{Tsybakov2009}] \label{lem:nonparametric_regression}
If $h = h_m = \alpha m^{-\frac{1}{2\lambda + 3}}$ for some $\alpha > 0$, the following estimation error bound holds for $\beta \in \Gamma^K$:
\begin{align}
\sup_{x \in \mathcal{X}} \mathbb{E}\left[(\beta(x) - \hat{\beta}_m^{\textsf{LP}}(x))^2\right] = O\left(m^{-\frac{2 (\lambda+1)}{2\lambda+3}}\right). \label{eq:nonpara_error}
\end{align}
\end{lem}
For further results in nonparametric regression, see \cite{Tsybakov2009, GyorfiKKW2002}.

Like for noiseless samples, the following theorem shows that the average loss $\bar{\Delta}_{\textsf{E}}(R)$ incurred due to sampled knowledge about $\beta$ is asymptotically negligible.
\begin{thm} \label{thm:delta_E_exp}
For $R \ge 0$,
\begin{align*}
\bar{\Delta}_{\textsf{E}}(R) = O\left( m^{-\frac{\lambda+1}{2\lambda+3}} \right).
\end{align*}
\end{thm}
\begin{IEEEproof}
First note that due to the Jensen's inequality,
\begin{align*}
&~ \sup_{x \in \mathcal{X}} \left( \mathbb{E}[|\beta(x) - \hat{\beta}_m^{\textsf{LP}}(x)|]\right)^2 \\
\le&~ \sup_{x \in \mathcal{X}} \mathbb{E} \left[ (\beta(x) - \hat{\beta}_m^{\textsf{LP}}(x))^2 \right] = O\left( m^{-\frac{2 (\lambda+1)}{2\lambda+3}} \right),
\end{align*}
which implies 
\begin{align}
\mathbb{E}[|\beta(x) - \hat{\beta}_m^{\textsf{LP}}(x)|] = O\left(m^{-\frac{\lambda+1}{2\lambda+3}} \right) ~~ \forall x \in \mathcal{X}. \label{eq:regression_bound}
\end{align}
Now fix $P_X$ so that rate $R = I(X;Y)$ is also fixed. The expectation in \eqref{eq:nonpara_error} is over the sampling noise distribution,
\begin{align*}
&~ \mathbb{E}_Z \left[ \left| \mathbb{E}_{X} [\beta(X)] - \mathbb{E}_{X} [\hat{\beta}_m^{\textsf{LP}}(X)] \right| \right] \\
\le&~ \mathbb{E}_Z \left[ \mathbb{E}_{X} [|\beta(X) - \hat{\beta}_m^{\textsf{LP}}(X)|] \right] \\
=&~ \mathbb{E}_X \left[ \mathbb{E}_{Z} [|\beta(X) - \hat{\beta}_m^{\textsf{LP}}(X)|] \right] \\
\stackrel{(a)}{\le}&~ \mathbb{E}_X \left[O\left(m^{-\frac{\lambda+1}{2\lambda+3}}\right)\right] \stackrel{(b)}{=} O\left(m^{-\frac{\lambda+1}{2\lambda+3}}\right),
\end{align*}
where (a) follows from \eqref{eq:regression_bound} and (b) follows since \eqref{eq:regression_bound} holds for every $x$. By the same argument as in the proof of Thm.~\ref{thm:delta_E_UB}, we know that
\begin{align*}
\bar{\Delta}_{\textsf{E}}(R; \beta, \hat{\beta}_m^{\textsf{LP}}) = \mathbb{E}_Z \left[ |B_{\beta}(R) - B_{\hat{\beta}_m^{\textsf{LP}}}(R)| \right] = O\left(m^{-\frac{\lambda+1}{2\lambda+3}}\right),
\end{align*}
which does not depend on $\beta$.

As $\beta \in \Gamma^K$, $R \ge 0$ are arbitrary, and the local polynomial estimator is a particular choice of estimator, taking the infimum over all estimators implies $\bar{\Delta}_{\textsf{E}}(R) \le \bar{\Delta}_{\textsf{E}}(R; \beta, \hat{\beta}_m^{\textsf{LP}}) = O\left(m^{-\frac{\lambda+1}{2\lambda+3}}\right)$.
\end{IEEEproof}

Paralleling arguments for noiseless samples, the information loss can be also specified.
\begin{cor} \label{cor:delta_I_exp}
The following are true:
\begin{enumerate}
\item For $B \in [0,B_{\textsf{max}})$, $\bar{\Delta}_{\textsf{I}}(B;\beta, \hat{\beta}_m) \to 0 ~~ \textrm{as } m \to \infty$.

\item Suppose the channel yields $M$-Lipschitz continuous $C_{\beta}(B)$ on $(0, B_{\textsf{max}})$. Then, $\bar{\Delta}_{\textsf{I}}(B;\beta, \hat{\beta}_m) = O\left(m^{-\frac{\lambda+1}{2\lambda+3}}\right)$ for all $B \in [0, B_{\textsf{max}})$.

\item There is a a pair of harvesting function and channel for which $\bar{\Delta}_{\textsf{I}}(B_{\textsf{max}};\beta, \hat{\beta}_m) = C_{\textsf{max}}$.
\end{enumerate}
\end{cor}
Proofs are basically the same as the proofs of Thm.~\ref{thm:delta_I} and Cor.~\ref{cor:delta_I_Lip}, so omitted.

\section{Sampling Loss in SIET Multicast} \label{sec:multicast}

Now we investigate the multicast setting in Fig.~\ref{fig:multicast}. Consider a single transmitter (i.e., access point) and $L$ receiver nodes. The transmitter sends a signal $X^n$ which conveys not only a common message $W$, but also energy to operate each node. These nodes observe $(Y^{(\ell)})^n$ through individual channels and have their own harvesting functions $\beta^{(\ell)} \in \Gamma^{K},  \ell = 1, \ldots, L$ and energy requirements $B^{(\ell)}$, which are not necessarily identical since physical devices may be different. As before, we are limited in knowing the harvesting functions only at sample points either in the absence or presence of noise.

The next proposition states the capacity-energy tradeoff for the SIET multicast problem with full knowledge of harvesting functions \cite{WuTVM2018_arXiv}. Here, superscript $(\textsf{MC})$ explicitly denotes that it is a multicast quantity. For notational simplicity, we use vector notations
\begin{align*}
\mathbf{B} &= [B^{(1)}, \ldots, B^{(L)}], \\
\bm{\beta} &= [\beta^{(1)}, \ldots, \beta^{(L)}], \\
\bm{\hat{\beta}}_m &= [\hat{\beta}_m^{(1)}, \ldots, \hat{\beta}_m^{(L)}], \\
\bm{\Gamma}(\bm{\beta}, m) &= [\Gamma(\beta^{(1)}, m), \ldots, \Gamma(\beta^{(L)}, m)].
\end{align*}

\begin{prop}[Thm.~1 in \cite{WuTVM2018_arXiv}] \label{prop:multicast_cap}
For $L$-user SIET multicast, the capacity-energy function is given by
\begin{align*}
C_{\bm{\beta}}^{(\textsf{MC})}(\mathbf{B}) = \max_{\substack{P_X: \forall \ell \\ \mathbb{E}[\beta^{(\ell)}(X)] \ge B^{(\ell)}} } \min_{1 \le \ell \le L} I(X;Y^{(\ell)}).
\end{align*}
\end{prop}
Also like \eqref{eq:C_F}, it is easy to extend to the set of possible harvesting functions.
\begin{align*}
C_{\bm{\Gamma}(\bm{\beta}, m)}(\mathbf{B}) = \max_{\substack{P_X: \forall \ell \\ \mathbb{E}[\beta^{(\ell)}(X)] \ge B^{(\ell)} \\ \forall \hat{\beta}^{(\ell)} \in \Gamma(\beta^{(\ell)},m)}} \min_{1 \le \ell \le L} I(X;Y^{(\ell)}).
\end{align*}

Let $B_{\bm{\beta}}^{(\ell)}(R), B_{\bm{\Gamma}(\bm{\beta}, m)}^{(\ell)}(R)$ be the amounts of energy delivered to $\ell$th node using the rate $R$ codebook designed for $\bm{\beta}$ and $\bm{\Gamma}(\bm{\beta}, m)$, respectively, that is,
\begin{align*}
B_{\bm{\beta}}^{(\ell)}(R) &= \max_{\substack{P_X: \forall \ell \\ I(X;Y^{(\ell)}) \ge R}} \mathbb{E}[\beta^{(\ell)}(X)] \\
B_{\bm{\Gamma}(\bm{\beta},m)}^{(\ell)}(R) &= \max_{\substack{P_X: \forall \ell \\ I(X;Y^{(\ell)}) \ge R}} \min_{\hat{\beta}_m \in \Gamma(\beta^{(\ell)}, m)} \mathbb{E}[\hat{\beta}_m(X)]
\end{align*}

Hence, sampling losses \eqref{eq:delta_i}--\eqref{eq:delta_e_exp} defined for the point-to-point case extend to multicast as follows. Note that $\Delta_{\textsf{E}}^{(\textsf{MC})}(R), \Delta_{\textsf{I}}^{(\textsf{MC})}(\mathbf{B})$ are for noiseless samples and $\bar{\Delta}_{\textsf{E}}^{(\textsf{MC})}(R), \bar{\Delta}_{\textsf{I}}^{(\textsf{MC})}(\mathbf{B})$ are for noisy samples.
\begin{align*}
&\Delta_{\textsf{E}}^{(\textsf{MC})}(R) = \sup_{\beta^{(\ell)} \in \Gamma^K} \max_{1 \le \ell \le L} B_{\bm{\beta}}^{(\ell)}(R) - B_{\bm{\Gamma}(\bm{\beta}, m)}^{(\ell)}(R), \\
&\Delta_{\textsf{I}}^{(\textsf{MC})}(\mathbf{B}; \bm{\beta}, \bm{\Gamma}(\bm{\beta},m)) = C_{\bm{\beta}}^{(L)}(\mathbf{B}) - C_{\bm{\Gamma}(\bm{\beta},m)}^{(L)}(\mathbf{B}), \\
&\bar{\Delta}_{\textsf{E}}^{(\textsf{MC})}(R) = \inf_{\hat{\beta}_m^{(\ell)}} \sup_{\beta^{(\ell)} \in \Gamma^K} \max_{1 \le \ell \le L} \mathbb{E}\left[ |B_{\bm{\beta}}^{(\ell)}(R) - B_{\bm{\hat{\beta}}_m}^{(\ell)}(R)| \right], \\
&\bar{\Delta}_{\textsf{I}}^{(\textsf{MC})}(\mathbf{B}; \bm{\beta}, \bm{\hat{\beta}}_m) = \mathbb{E}\left[ |C_{\bm{\beta}}^{(L)}(\mathbf{B}) - C_{\bm{\hat{\beta}}_m}^{(L)}(\mathbf{B})| \right].
\end{align*}
Note that $\Delta_{\textsf{I}}^{(\textsf{MC})}(\mathbf{B}), \bar{\Delta}_{\textsf{I}}^{(\textsf{MC})}(\mathbf{B})$ do not have maximum over $\ell$ because all nodes receive the same information in multicast.

\begin{thm}[Noiseless samples] \label{thm:multicast_noiseless}
The asymptotic bounds in Thms.~\ref{thm:delta_E_UB}, \ref{thm:delta_I} and Cor.~\ref{cor:delta_I_Lip} hold for multicast when samples are noiseless, that is:

\begin{enumerate}
	\item $\Delta_{\textsf{E}}^{(\textsf{MC})}(R) = O(m^{-\lambda})$.
	
	\item $\Delta_{\textsf{I}}^{(\textsf{MC})}(\mathbf{B};\bm{\beta}, \bm{\Gamma}(\bm{\beta},m)) \to 0$ as $m \to \infty$ if $B^{(\ell)} \in [0, B_{\textsf{max}}^{(\ell)})$ for all $\ell$.
	
	\item Letting $C_{\textsf{max}}^{(\textsf{MC})} := \max_{P_X} \min_{1 \le \ell \le L} I(X;Y^{(\ell)})$, there exists a channel such that $\Delta_{\textsf{I}}^{(\textsf{MC})}(\mathbf{B};\bm{\beta}, \bm{\Gamma}(\bm{\beta},m)) = C_{\textsf{max}}^{(\textsf{MC})}$ if some $B^{(\ell)} = B_{\textsf{max}}^{(\ell)}$.
	
	\item Suppose $C_{\bm{\beta}}^{(\textsf{MC})}(\mathbf{B})$ is $M$-Lipschitz with $\ell_q$ norm, where $1 \le q \le \infty$, that is,
	\begin{align*}
		|C_{\bm{\beta}}(\mathbf{B}_1) - C_{\bm{\beta}}(\mathbf{B}_2)| \le M || \mathbf{B}_1 - \mathbf{B}_2 ||_q.
	\end{align*}
	Then, $\Delta_{\textsf{I}}^{(\textsf{MC})}(\mathbf{B};\bm{\beta}, \bm{\Gamma}(\bm{\beta},m)) = O(m^{-\lambda})$.
\end{enumerate}
\end{thm}

\begin{thm}[Noisy samples] \label{thm:multicast_noisy}
The asymptotic bounds in Thm.~\ref{thm:delta_E_exp} and Cor.~\ref{cor:delta_I_exp} also hold for multicast when samples are noisy, that is,
\begin{enumerate}
\item $\bar{\Delta}_{\textsf{E}}^{(\textsf{MC})}(R) = O\left(m^{-\frac{\lambda+1}{2\lambda+3}}\right)$.

\item $\bar{\Delta}_{\textsf{I}}^{(\textsf{MC})}(\mathbf{B}; \bm{\beta}, \bm{\hat{\beta}}_m) \to 0$ as $m \to \infty$ if $B^{(\ell)} \in [0, B_{\textsf{max}}^{(\ell)})$ for all $\ell$.

\item Suppose $C_{\bm{\beta}}^{(\textsf{MC})}(\mathbf{B})$ is $M$-Lipschitz with $\ell_q$ norm, where $1 \le q \le \infty$, that is,
\begin{align*}
|C_{\bm{\beta}}(\mathbf{B}_1) - C_{\bm{\beta}}(\mathbf{B}_2)| \le M || \mathbf{B}_1 - \mathbf{B}_2 ||_q.
\end{align*}
Then, $\bar{\Delta}_{\textsf{I}}^{(\textsf{MC})}(\mathbf{B}; \bm{\beta}, \bm{\hat{\beta}}_m) = O\left(m^{-\frac{\lambda+1}{2\lambda+3}}\right)$.
\end{enumerate}
\end{thm}

We omits proofs of both theorems since proof techniques follow the point-to-point proofs.

\section{End-to-End Communication with Samples} \label{sec:JSCC_approx}
\begin{figure}[t]
	\centering
	\includegraphics[width=2.5in]{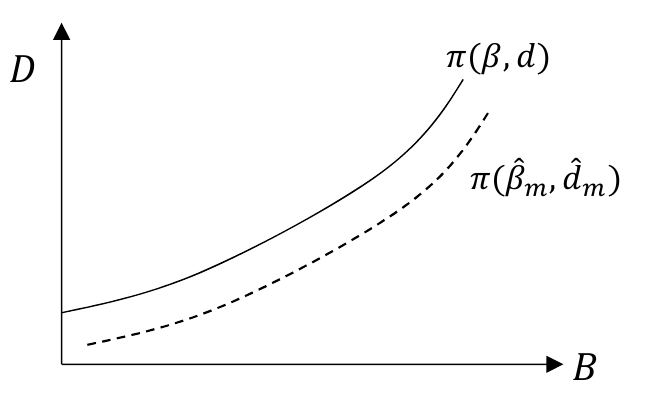}
	\caption{An illustration of the energy-distortion tradeoff. The dotted line is for the estimated harvesting and distortion functions.}
	\label{fig:energy_distortion_tradeoff}
\end{figure}

Consider the end-to-end information transmission problem in the SIET framework, which consists of source and channel components. The first is a source-distortion pair $(P_S, d)$, where a source sequence $\{S_i\}$ is drawn from $P_S$ on $\mathcal{S}$, and a non-negative distortion measure $d:\mathcal{S} \times \hat{\mathcal{S}} \to \mathbb{R}_+$ is given. The second is a channel-harvesting pair $(P_{Y|X}, \beta)$, where $\beta$ is a non-negative energy harvesting function. Note that unlike the standard problem where there is a channel cost \textit{constraint}, here there is an energy \textit{requirement}.

In the end-to-end transmission problem, the goal is to minimize distortion $D$ between the two terminals, but also maximize energy transmission $B$. That is, the goal is to find the best energy-distortion pair $(B,D)$ such that $\mathbb{E}[d(S,\hat{S})] \le D$ and $\mathbb{E}[\beta(X)] \ge B$. For given harvesting and distortion functions $(\beta, d)$, we can define the optimal $(B,D)$ tradeoff curve (perhaps degenerate), $\pi(\beta, d)$, as follows \cite{Gastpar2003}.
\begin{defi} \label{def:optiomal_JSCC}
	The curve $\pi(\beta, d)$ is said to be optimal if every $(B, D) \in \pi(\beta, d)$ satisfies both of the followings.
	\begin{enumerate}
		\item $D$ cannot be decreased without decreasing $B$.
		\item $B$ cannot be increased without increasing $D$.
	\end{enumerate}
\end{defi}

A typical $\pi(\beta, d)$ curve is illustrated in Fig.~\ref{fig:energy_distortion_tradeoff}. It is continuous, monotone increasing, and convex if non-degenerate. The monotonicity is due to Def.~\ref{def:optiomal_JSCC}. In addition, if it is non-convex, the curve can be improved by time-sharing so we can conclude it is convex. Continuity follows from convexity.

In place of full knowledge of $(d, \beta)$, we only have samples for both distortion and harvesting functions so we have $(\hat{d}_m, \hat{\beta}_m)$. Informally, $(\hat{d}_m, \hat{\beta}_m)$ is close to the true pair when the number of samples is large. Analogous to our main result in Sec.~\ref{sec:SIET approx} for the SIET channel coding problem, the source coding problem with sampled distortion measure was studied by Niesen, et al.~\cite{NiesenSW2006} who showed that the distortion loss vanishes as the number of samples increases. See the Appendix for detailed problem setting and results with its extension to noisy samples. Further, we have shown that designing codebooks as if $(\hat{\beta}_m, \hat{d}_m)$ are the true functions is nearly optimal for noiseless and noisy cases. Hence, the question that naturally follows is whether $\pi(\hat{\beta}_m, \hat{d}_m)$ is also close to $\pi(\beta, d)$.

For two optimal tradeoff curves $\pi(\beta, d), \pi(\hat{\beta}_m, \hat{d}_m)$, let us define loss incurred by sampling. Let $\Pi_{\pi}(B,D)$ be the projection of $(B,D)$ onto curve $\pi$ under $\ell_1$ distance; when there are several projection points, pick any one arbitrarily. Then we define two component losses for noiseless and noisy samples, respectively, as\footnote{Note that $\Delta(\beta, d, \hat{\beta}_m, \hat{d}_m), \bar{\Delta}(\beta, d, \hat{\beta}_m, \hat{d}_m)$ are well-defined even when $\pi(\beta, d)$ or $\pi(\hat{\beta}_m, \hat{d}_m)$ is degenerate.}
\begin{align*}
&~ \Delta(\beta, d, \hat{\beta}_m, \hat{d}_m) \\
=&~ \sup_{(B', D') \in \pi(\hat{\beta}_m, \hat{d}_m)} ||(B', D') - \Pi_{\pi(\beta, d)}(B',D')||_1, \\
&~ \bar{\Delta}(\beta, d, \hat{\beta}_m, \hat{d}_m) \\
=&~ \sup_{(B', D') \in \pi(\hat{\beta}_m, \hat{d}_m)} \mathbb{E}\left[ ||(B', D') - \Pi_{\pi(\beta, d)}(B',D')||_1 \right].
\end{align*}

By definition, $\Delta, \bar{\Delta}$ are the maximal possible losses from the true optimal curve when we design optimal end-to-end transmission as if $(\hat{\beta}_m, \hat{d}_m)$ is the true harvesting and distortion function pair.\footnote{Also we can consider the other direction of projection, which is projection from $\pi(\beta, d)$ onto $\pi(\hat{\beta}_m, \hat{d}_m)$. But, since what we want to know is how close our estimation is to the true one, this makes less sense in practice.} By Shannon's separation theorem \cite{Shannon1959}, any operating point $(B,D)$ in $\pi$ can be attained by a separately designed pair of good source and channel codes. Moreover, distortion loss in source coding and harvesting loss in channel coding due to sampling vanish by results in \cite{NiesenSW2006} (restated in Appendix) and Sec.~\ref{sec:SIET approx}.  Thus one might conjecture that a system design based on $(\hat{\beta}_m, \hat{d}_m)$ is nearly optimal, i.e., $\Delta, \bar{\Delta} \to 0$ as $m \to \infty$. This is partially true with additional restricion on harvesting and distortion functions. The following theorem formally shows it.

\begin{figure}
	\centering
	\includegraphics[width=3.5in]{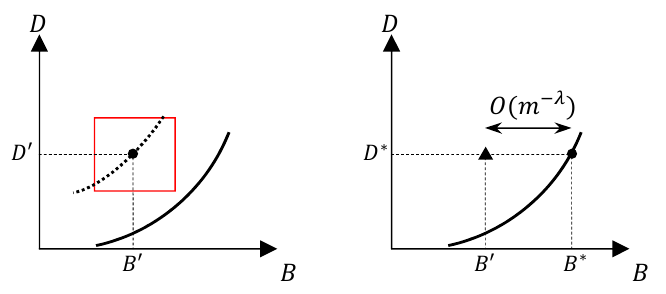}
	\caption{The proof of Thm.~\ref{thm:JSCC_pf}. Solid curves and dotted curves denote $\pi(\beta, d)$ and $\pi(\hat{\beta}_m, \hat{d}_m)$, respectively. The left illustrates that there is no point in the $\ell_1$-ball centered at $(B',D')$, drawn in red. The right illustrates that the channel codebook at $B^*$ performs $B'$, marked as triangle, under $\hat{\beta}_m$.}
	\label{fig:JSCC_pf}
\end{figure}

\begin{figure}
	\centering
	\includegraphics[width=3.5in]{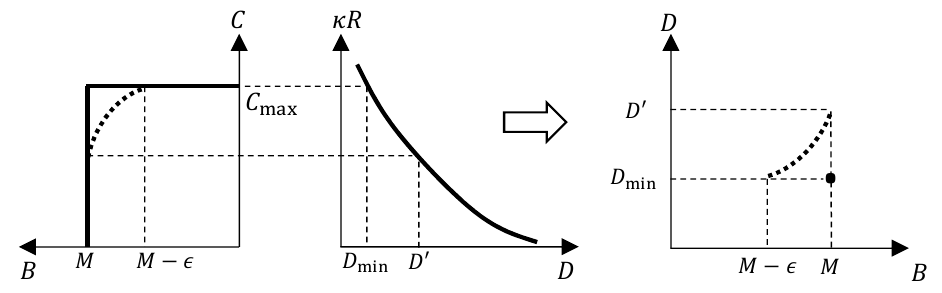}
	\caption{An example in the proof of Thm.~\ref{thm:JSCC_counter_ex}. Solid curves denote true quantities $C_{\beta}(B), R_{d}(D), \pi(\beta, d)$ and dotted curves denote quantities for estimated functions. Note that $d = \hat{d}_m$ and $\pi(\beta, d)$ is degenerate. $\kappa = \frac{k_1}{k_2}$ indicates the ratio that $k_1$ source symbols are mapped to $k_2$ channel symbols.}
	\label{fig:JSCC_counter_ex}
\end{figure}

\begin{thm} \label{thm:JSCC_pf}
Define two sets,
\begin{align*}
\mathcal{B} &:=\{\beta \in \Gamma^K: C_{\beta}(B) \textrm{ is Lipschitz over all } B \ge 0 \}, \\
\mathcal{D} &:=\{d(\cdot, \hat{s}) \in \Gamma^K ~ \forall \hat{s}: R_{d}(D) \textrm{ is Lipschitz over all } D \ge 0 \},
\end{align*}
and two minimax losses
\begin{align*}
\Delta := \inf_{\hat{\beta}_m, \hat{d}_m} \sup_{\beta \in \mathcal{B}, d \in \mathcal{D}} \Delta(\beta, d, \hat{\beta}_m, \hat{d}_m), \\
\bar{\Delta} := \inf_{\hat{\beta}_m, \hat{d}_m} \sup_{\beta \in \mathcal{B}, d \in \mathcal{D}} \bar{\Delta}(\beta, d, \hat{\beta}_m, \hat{d}_m).
\end{align*}
Then, $\Delta = O(m^{-\lambda})$ and $\bar{\Delta} = O\left( m^{-\frac{\lambda+1}{2\lambda + 3}} \right)$.
\end{thm}
\begin{IEEEproof}
Consider $(\hat{\beta}_m, \hat{d}_m)$ are estimated by the spline method for noiseless samples and by the local polynomial regression for noisy samples, i.e., $(\hat{\beta}_m,\hat{d}_m) = (\hat{\beta}_m^{\textsf{SP}}, \hat{d}_m^{\textsf{SP}})$ for noiseless and $(\hat{\beta}_m,\hat{d}_m) = (\hat{\beta}_m^{\textsf{LP}}, \hat{d}_m^{\textsf{LP}})$ for noisy samples. Let us only focus on noiseless samples. Proof will be shown by contradiction: suppose that there exists $(B',D') \in \pi(\hat{\beta}_m^{\textsf{SP}}, \hat{d}_m^{\textsf{SP}})$ such that the  $\ell_1$-balls centered at $(B',D')$ with radius $O(m^{-\lambda})$ has no intersection with $\pi(\beta, d)$.

First consider the optimal codebook pair at $(B',D')$. Although the channel codebook is designed for $\hat{\beta}_m^{\textsf{SP}}$, actual harvested energy $B$ is also close to $B'$, i.e., $B = B'+O(m^{-\lambda})$. Similarly, the source codebook also achieves the actual distortion $D = D' + O(m^{-\lambda})$. Since these codebooks are suboptimal for the true $(\beta, d)$, there will be a point $(B^*,D^*)$ on $\pi(\beta, d)$ such that $B^* \ge B \ge B' - cm^{-\lambda}$ and $D^* \le D \le D'+cm^{-\lambda}$.

Pick a point $(B^*, D^*) \in \pi(\beta, d)$ such that $D^* = D'$. We know that this point exists from the Lipschitz continuity. From the assumption, we know that $B^*$ is outside of the $\ell_1$-ball, i.e., $B^* > B' + cm^{-\lambda}$. Consider the optimal codebook pair at $(B^*, D^*)$. From the first argument of the proof of Thm.~\ref{thm:delta_E_UB}, we know that the channel codebook delivers energy $B^*+O(m^{-\lambda})$ under harvesting function $\hat{\beta}_m^{\textsf{SP}}$. However, this codebook is definitely suboptimal for $\hat{\beta}_m^{\textsf{SP}}$, which means that $\pi(\hat{\beta}_m^{\textsf{SP}}, \hat{d}_m^{\textsf{SP}})$ has a point $(D', B')$ such that $B' > B^* - cm^{-\lambda}$. This implies $|B'-B^*| \le cm^{-\lambda}$, a contradiction. Therefore, $\Delta(\beta, d, \hat{\beta}_m^{\textsf{SP}}, \hat{d}_m^{\textsf{SP}}) =O(m^{-\lambda})$. Since the bound is independent of $(\beta, d)$ and $(\hat{\beta}_m^{\textsf{SP}}, \hat{d}_m^{\textsf{SP}})$ are specific reconstructions, we can further reduce the loss. Therefore, $\Delta = O(m^{-\lambda})$ holds. The argument is illustrated in Fig.~\ref{fig:JSCC_pf}.

For noisy sample, the arguments still hold with $(\hat{\beta}_m^{\textsf{LP}}, \hat{d}_m^{\textsf{LP}})$ so $\bar{\Delta} = O\left( m^{-\frac{\lambda+1}{2\lambda + 3}} \right)$.
\end{IEEEproof}

Despite the above theorem showing $\Delta, \bar{\Delta}$ converge to zero for $\mathcal{B}, \mathcal{D}$, the next theorem demonstrates its components $\Delta(\beta, d, \hat{\beta}_m, \hat{d}_m)$ and $\bar{\Delta}(\beta, d, \hat{\beta}_m, \hat{d}_m)$ could be arbitrary large unless $\beta \in \mathcal{B}, d \in \mathcal{D}$, even when $\hat{\beta}_m, \hat{d}_m$ are sufficiently accurate. It suggests the possibility that accurate reconstruction may not be enough to provide performance guarantee for end-to-end communication.

\begin{thm}\label{thm:JSCC_counter_ex}
	There exists a case where $\Delta(\beta, d, \hat{\beta}_m, \hat{d}_m), \bar{\Delta}(\beta, d, \hat{\beta}_m, \hat{d}_m)$ are bounded away from $0$ even when $m \to \infty$.
\end{thm}
\begin{IEEEproof}
	Consider an example with noiseless samples illustrated in Fig.~\ref{fig:JSCC_counter_ex}. For the source coding part, suppose the $R_d$ curve is strictly convex and assume that our estimate is perfect, i.e., $d = \hat{d}_m$ so that $R_d(D) = R_{\hat{d}_m}(D)$.
	
	For the channel part, suppose $\beta(x) = M$ for all $x \in \mathcal{X}$ for some constant $M$. Then, every $P_X$ is admissible with respect to energy requirement $M$ since every $P_X$ achieves $\mathbb{E}[\beta(X)] = M$. Let $C_{\textsf{max}} = \max_{P_X} I(X;Y)$ and $P_X^*$ be the unique capacity-achieving distribution which is non-vanishing everywhere on $\mathcal{X}$. By the separation theorem, this combination yields a degenerate JSCC curve $\pi(\beta, d) = (M, D_{\textsf{min}})$. $C_{\beta}(B), R_{d}(D), \pi(\beta, d)$ are illustrated with solid line.
	
	On the other hand, suppose our estimate is $\hat{\beta}_m(x) = M(1-f(x))$, where $f(x)$ is a small non-negative bumpy function such that $f(x_i) = 0$ only at every $x_i$. There are two end points in $C_{\hat{\beta}_m}$: One point is induced by $P_X^*$, which still achieves the best in information delivery, however, $\mathbb{E}_{P_X^*}[\hat{\beta}_m] = M-\epsilon$ for some $\epsilon > 0$. The other is by some discrete probability, that is, engineers design a codebook that only utilizes a finite number of points in $\mathcal{X}$, which is strictly suboptimal in information transmission. Since $\hat{\beta}_m(x) = M$ only at $x_i$, the transmitted energy is maximized when $P_X$ has only point masses on $x_i$, but such restriction on distribution incurs non-vanishing mutual information loss. Therefore resulting $\pi(\hat{\beta}_m, \hat{d}_m)$ is a convex curve connecting $(M-\epsilon, D_{\textsf{min}})$ and $(M,D')$. Therefore,
	\begin{align*}
	||(M, D') - \Pi_{\pi(\beta, d)}(M,D')||_1 = D'-D_{\textsf{min}},
	\end{align*}
	which is non-vanishing, so $\Delta(\beta, d, \hat{\beta}_m, \hat{d}_m)$ is also non-vanishing.
	
	The argument for $\bar{\Delta}(\beta, d, \hat{\beta}_m, \hat{d}_m)$ is immediate since $\bar{\Delta}(\beta, d, \hat{\beta}_m, \hat{d}_m) \ge \Delta(\beta, d, \hat{\beta}_m, \hat{d}_m)$.
\end{IEEEproof}

\section{Conclusion} \label{sec:conclusion}
We have studied performance loss in SIET due to experimentally-sampled harvesting functions. To our knowledge, this is the first study of how sampled knowledge of perhaps nonlinear and nonideal harvesting circuits affects SIET (or SWIPT). Energy loss and information loss are separately considered for noiseless and noisy samples, and extended to multicast setting. We show theoretical asymptotics for these losses that energy loss asymptotically vanishes as $O(m^{-\lambda})$ for noiseless samples and it is indeed asymptotically optimal under some technical conditions. For noisy samples, the speed of convergence in energy loss is lowered to $O(m^{-\frac{\lambda+1}{2\lambda+3}})$ due to noise in characterizing the harvesting circuit.

We also suggest spline and local polynomial reconstruction as practical reconstruction methods that attain the above asymptotics. B-spline (basis-spline) method requires $O(m)$ complexity \cite{ToraichiKSM1987} and the local polynomial estimator at each $x$ requires complexity at most polynomial in $m$ since \eqref{eq:local_poly} resulted from matrix algebra.

With regard to information loss, large number of samples does not always guarantee vanishing information loss. To get a vanishing information loss, a certain energy margin from $B_{\textsf{max}}$ needs to be guaranteed. Hence, it is necessary for system designers to set a sufficient energy transmission margin from $B_{\textsf{max}}$.

Another important problem is end-to-end information transmission. Motivated by \cite{NiesenSW2006}, which shows the optimal source code for a sampled distortion function is also near-optimal for the true distortion function, one might guess that Shannon's separation theorem would yield a combination of near-optimal source code and channel code that combine to be near-optimal in the energy-distortion tradeoff. It is true when further restriction is given on harvesting and distortion functions.

\section*{Appendix}
Let us restate the main result of \cite{NiesenSW2006}, which considers the lossy source coding problem with noiseless samples of the distortion function. The following assumptions are made on the source component. Suppose $\mathcal{S} = [0,1]$, $\hat{\mathcal{S}}$ is some discrete set, and $d(\cdot, \hat{s}) \in \Gamma^K$ for all $\hat{s} \in \hat{\mathcal{S}}$. For instance, $\mathcal{S}$ is a set of images, $\hat{\mathcal{S}}$ is a set of quantized images or labels of images, and $d(s, \hat{s})$ is human perception loss which is unknown. Like a harvesting function, only a finite number of evenly-spaced sample points of $d$ are known. In particular, for each $\hat{s} \in \hat{\mathcal{S}}$, $\{(s_i, d(s_i, \hat{s}))\}_{i=0}^{m-1}$ are given by experiment, where $s_i = \frac{i}{m-1}$. So $m \times |\hat{\mathcal{S}}|$ samples are given. In the case of noisy samples, $\{(s_i, d(s_i, \hat{s}) + Z_i )\}_{i=0}^{m-1}$ are given for each $\hat{s} \in \mathcal{S}$, where $Z_i$ is i.i.d.~additive noise with mean zero and variance $\sigma_2^2$. 

For a distortion function $f$ and a set of distortion functions $F$, \emph{rate-distortion} functions are defined as
\begin{align*}
R_{f}(D) &= \inf_{P_{\hat{S}|S}:\mathbb{E}[f(S,\hat{S})] \le D} I(S;\hat{S}), \\
R_{F}(D) &= \min_{P_{\hat{S}|S}: \mathbb{E}[f(S,\hat{S})] \le D ~ \forall f \in F } I(S;\hat{S}).
\end{align*}
\emph{Distortion-rate} functions are defined as
\begin{align*}
D_{f}(R) &= \min_{P_{\hat{S}|S}:I(S;\hat{S}) \le R} \mathbb{E}[f(S,\hat{S})], \\
D_{F}(R) &= \min_{P_{\hat{S}|S}:I(S;\hat{S}) \le R} \max_{f \in F} \mathbb{E}[f(S,\hat{S})].
\end{align*}
Then, the sampling loss in distortion for noiseless samples is defined as
\begin{align*}
\Delta_{\textsf{D}}(R) = \sup_{d \in \Gamma^K} D_{\Gamma(d, m)}(R) - D_d(R).
\end{align*}
For noisy samples, we can generalize the distortion loss to noisy samples, similarly to \eqref{eq:delta_e_exp}.
\begin{align*}
\bar{\Delta}_{\textsf{D}}(R) = \inf_{\hat{d}_m} \sup_{d \in \Gamma^K} \mathbb{E} \left[ |D_d(R) - D_{\hat{d}_m}(R)| \right],
\end{align*}
where $\hat{d}_m$ is the estimate of the distortion function. Then, we have the following distortion bound for noiseless samples.
\begin{lem}[Thm.~1 in \cite{NiesenSW2006}] \label{lem:delta_D}
	If $P_S(s) < c ~~ \forall s \in \mathcal{S}$ with some constant $c$,
	\begin{align*}
	\Delta_{\textsf{D}}(R) = O(m^{-\lambda}).
	\end{align*} 
\end{lem}

We generalize to the noisy samples case as follows.
\begin{lem} \label{lem:delta_D_exp}
	If $P_S(s) < c ~~ \forall s \in \mathcal{S}$ with some constant $c$,
	\begin{align*}
	\bar{\Delta}_{\textsf{D}}(R) = O\left(m^{-\frac{\lambda+1}{2\lambda+3}} \right).
	\end{align*}
\end{lem}
\begin{IEEEproof}
Pick an arbitrary compression kernel $P_{\hat{S}|S}$. Then, rate $R = I(S;\hat{S})$ is also fixed. For given $(d, \hat{d}_m^{\textsf{LP}})$, noting that the expectation is over the noise distribution,
\begin{align}
&~ \mathbb{E}_Z \left[ \left| \mathbb{E}_{S, \hat{S}}[d(S, \hat{S})] - \mathbb{E}_{S, \hat{S}}[\hat{d}_m^{\textsf{LP}}(S, \hat{S})] \right| \right] \nonumber \\
\le&~ \mathbb{E}_Z \left[ \mathbb{E}_{S, \hat{S}}[ |d(S, \hat{S}) - \hat{d}_m^{\textsf{LP}}(S, \hat{S})| ] \right] \nonumber \\
=&~ \mathbb{E}_{S, \hat{S}} \left[ \mathbb{E}_Z [ |d(S, \hat{S}) - \hat{d}_m^{\textsf{LP}}(S, \hat{S})| ] \right] \nonumber \\
=&~ \sum_{\hat{s} \in \hat{\mathcal{S}}} \int_{\mathcal{S}} P_S(s) P_{\hat{S}|S}(\hat{s}|s) \mathbb{E}_Z [ |d(s, \hat{s}) - \hat{d}_m^{\textsf{LP}}(s, \hat{s})| ] ds. \label{eq:bound1}
\end{align}
As $P_S(s) \le c$ and $P_{\hat{S}|S}(\hat{s}|s) \le 1$ for all $\hat{s} \in \hat{\mathcal{S}}$, \eqref{eq:bound1} can be further bounded.
\begin{align*}
\eqref{eq:bound1} ~ &\le c \sum_{\hat{s} \in \hat{\mathcal{S}}} \int_{\mathcal{S}} \mathbb{E}_Z [ |d(s, \hat{s}) - \hat{d}_m^{\textsf{LP}}(s, \hat{s})| ] ds \\
&\le c' \sum_{\hat{s} \in \hat{\mathcal{S}}} \int_{\mathcal{S}} m^{-\frac{\lambda+1}{2\lambda + 3}} ds \\
&= c' |\hat{\mathcal{S}}| m^{-\frac{\lambda+1}{2\lambda + 3}} = O\left(m^{-\frac{\lambda+1}{2\lambda+3}} \right),
\end{align*}
where the last inequality follows from the local polynomial estimator in Lem.~\ref{lem:nonparametric_regression}. By the same argument as in the proof of Thm.~\ref{thm:delta_E_UB} we have
\begin{align*}
\mathbb{E}_Z \left[ |D_d(R) - D_{\hat{d}_m^{\textsf{LP}}}(R)| \right] = O\left(m^{-\frac{\lambda+1}{2\lambda + 3}}\right).
\end{align*}
Since the bound does not depend on the choice of $d(\cdot, \hat{s}) \in \Gamma^K$, infimum over estimators only further improves the loss of the local polynomial estimator,
\begin{align*}
\bar{\Delta}_{\textsf{D}}(R) = \inf_{\hat{d}_m} \sup_{d \in \Gamma^K} \mathbb{E} \left[ |D_d(R) - D_{\hat{d}_m}(R)| \right] = O\left(m^{-\frac{\lambda+1}{2\lambda + 3}}\right).
\end{align*}
\end{IEEEproof}

\section*{Acknowledgment}
We thank the anonymous reviewers for their suggestions which greatly improved and clarified this paper.


\begin{thebibliography}{99}
	\bibitem{Varshney2008}
	L.~R. Varshney, ``Transporting information and energy simultaneously,'' in
	\emph{Proc. 2008 IEEE Int. Symp. Inf. Theory}, Jul. 2008, pp. 1612--1616.
	
	\bibitem{Varshney2012}
	------, ``On energy/information cross-layer architectures,'' in \emph{Proc.
		2012 IEEE Int. Symp. Inf. Theory}, Jul. 2012, pp. 1361--1365.
	
	\bibitem{ZhangMH2015}
	R.~Zhang, R.~G. Maunder, and L.~Hanzo, ``Wireless information and power
	transfer: From scientific hypothesis to engineering practice,'' \emph{{IEEE}
		Commun. Mag.}, vol.~53, no.~8, pp. 99--105, Aug. 2015.
	
	\bibitem{ClerckxZSNKP2019}
	B.~Clerckx, R.~Zhang, R.~Schober, D.~W.~K. Ng, D.~I. Kim, and H.~V. Poor,
	``Fundamentals of wireless information and power transfer: From {RF} energy
	harvester models to signal and system designs,'' \emph{{IEEE} J. Sel. Areas
		Commun.}, vol.~37, no.~1, pp. 4--33, Jan. 2019.
	
	\bibitem{ZhouZH2013}
	X.~Zhou, R.~Zhang, and C.~K. Ho, ``Wireless information and power transfer:
	Architecture design and rate-energy tradeoff,'' \emph{{IEEE} Trans. Commun.},
	vol.~61, no.~11, pp. 4754--4767, Nov. 2013.
	
	\bibitem{SoyataCHsoyata2016}
	T.~Soyata, L.~Copeland, and W.~Heinzelman, ``{RF} energy harvesting for
	embedded systems: A survey of tradeoffs and methodology,'' \emph{{IEEE}
		Circuits Syst. Mag.}, vol.~16, no.~1, pp. 22--57, 2016.
	
	\bibitem{ValentaD2014}
	C.~R. Valenta and G.~D. Durgin, ``Harvesting wireless power: Survey of
	energy-harvester conversion efficiency in far-field, wireless power transfer
	systems,'' \emph{{IEEE} Microw. Mag.}, vol.~15, no.~4, pp. 108--120, Jun.
	2014.
	
	\bibitem{BoshkovskaNZS2015}
	E.~Boshkovska, D.~W.~K. Ng, N.~Zlatanov, and R.~Schober, ``Practical non-linear
	energy harvesting model and resource allocation for {SWIPT} systems,''
	\emph{{IEEE} Commun. Lett.}, vol.~19, no.~12, pp. 2082--2085, Dec. 2015.
	
	\bibitem{KangKK2018}
	J.-M. Kang, I.-M. Kim, and D.~I. Kim, ``Wireless information and power
	transfer: Rate-energy tradeoff for nonlinear energy harvesting,''
	\emph{{IEEE} Commun. Lett.}, vol.~17, no.~3, pp. 1966--1981, Mar. 2018.
	
	\bibitem{VarastehRC2017}
	M.~Varasteh, B.~Rassouli, and B.~Clerckx, ``Wireless information and power
	transfer over an {AWGN} channel: Nonlinearity and asymmetric {G}aussian
	signaling,'' in \emph{Proc. IEEE Inf. Theory Workshop (ITW'17)}, Nov. 2017,
	pp. 181--185.
	
	\bibitem{LeMF2008}
	T.~Le, K.~Mayaram, and T.~Fiez, ``Efficient far-field radio frequency energy
	harvesting for passively powered sensor networks,'' \emph{{IEEE} J.
		Solid-State Circuits}, vol.~43, no.~5, pp. 1287--1302, May 2008.
	
	\bibitem{StoopmanKVPS2013}
	M.~Stoopman, S.~Keyrouz, H.~J. Visser, K.~Philips, and W.~A. Serdijn, ``A
	self-calibrating {RF} energy harvester generating $1${V} at $-26.3$ {dBm},''
	in \emph{2013 Symp. VLSI Circuits Dig. Tech. Pap.}, Jun. 2013, pp.
	C226--C227.
	
	\bibitem{StoopmanKVPS2014}
	------, ``Co-design of a {CMOS} rectifier and small loop antenna for highly
	sensitive {RF} energy harvesters,'' \emph{{IEEE} J. Solid-State Circuits},
	vol.~49, no.~3, pp. 622--634, Mar. 2014.
	
	\bibitem{SamplePSS2013}
	A.~P. Sample, A.~N. Parks, S.~Southwood, and J.~R. Smith, ``Wireless ambient
	radio power,'' in \emph{Wirelessly Powered Sensor Networks and Computational
		{RFID}}, J.~R. Smith, Ed.\hskip 1em plus 0.5em minus 0.4em\relax New York:
	Springer, 2013, pp. 223--234.
	
	\bibitem{BaroudiQM2015}
	U.~Baroudi, A.~Qureshi, and S.~Mekid, ``Characterization and modeling of
	received signal strength and charging time for wireless energy transfer,''
	\emph{Adv. Electr. Eng.}, vol. 2015, 2015.
	
	\bibitem{Zhang_nature2019}
	X.~Zhang, J.~Grajal, J.~L. Vazquez-Roy, U.~Radhakrishna, X.~Wang, W.~Chern,
	L.~Zhou, Y.~Lin, P.-C. Shen, X.~Ji, X.~Ling, A.~Zubair, Y.~Zhang, H.~Wang,
	M.~Dubey, J.~Kong, M.~Dresselhaus, and T.~Palacios, ``Two-dimensional
	$\textrm{MoS}_2$-enabled flexible rectenna for {Wi-Fi}-band wireless energy
	harvesting,'' \emph{Nature}, Jan. 2019.
	
	\bibitem{Clerckx2018}
	B.~Clerckx, ``Wireless information and power transfer: Nonlinearity, waveform
	design, and rate-energy tradeoff,'' \emph{{IEEE} Trans. Signal Process.},
	vol.~66, no.~4, pp. 847--862, Feb. 2018.
	
	\bibitem{NintanavongsaMLR2012}
	P.~Nintanavongsa, U.~Muncuk, D.~R. Lewis, and K.~Roy~Chowdhury, ``Design
	optimization and implementation for {RF} energy harvesting circuits,''
	\emph{IEEE J. Emerg. Sel. Topics Circuits Syst.}, vol.~2, no.~1, pp. 24--33,
	Mar. 2012.
	
	\bibitem{HsiaoK1997}
	G.~C. Hsiao and R.~E. Kleinman, ``Mathematical foundations for error estimation
	in numerical solutions of integral equations in electromagnetics,''
	\emph{{IEEE} Trans. Antennas Propag.}, vol.~45, no.~3, pp. 316--328, Mar.
	1997.
	
	\bibitem{UnserD1997}
	M.~Unser and I.~Daubechies, ``On the approximation power of convolution-based
	least squares versus interpolation,'' \emph{{IEEE} Trans. Signal Process.},
	vol.~45, no.~7, pp. 1697--1711, Jul. 1997.
	
	\bibitem{Tsybakov2009}
	A.~B. Tsybakov, \emph{Introduction to Nonparametric Estimation}.\hskip 1em plus
	0.5em minus 0.4em\relax New York: Springer-Verlag, 2009.
	
	\bibitem{FouladgarS2012}
	A.~M. Fouladgar and O.~Simeone, ``On the transfer of information and energy in
	multi-user systems,'' \emph{{IEEE} Commun. Lett.}, vol.~16, no.~11, pp.
	1733--1736, Nov. 2012.
	
	\bibitem{AmorPKP2017}
	S.~B. Amor, S.~M. Perlaza, I.~Krikidis, and H.~V. Poor, ``Feedback enhances
	simultaneous wireless information and energy transmission in multiple access
	channels,'' \emph{{IEEE} Trans. Inf. Theory}, vol.~63, no.~8, pp. 5244--5265,
	Aug. 2017.
	
	\bibitem{TallaKRNSG2017}
	V.~Talla, B.~Kellogg, B.~Ransford, S.~Naderiparizi, J.~R. Smith, and
	S.~Gollakota, ``Powering the next billion devices with {Wi-Fi},''
	\emph{Commun. ACM}, vol.~60, no.~3, pp. 83--91, Mar. 2017.
	
	\bibitem{NiesenSW2006}
	U.~Niesen, D.~Shah, and G.~Wornell, ``Sampling distortion measures,'' in
	\emph{Proc. 44th Annu. Allerton Conf. Commun. Control Comput.}, Sep. 2006.
	
	\bibitem{Varshney2010}
	L.~R. Varshney, ``Unreliable and resource-constrained decoding,'' Ph.D.~thesis,
	Massachusetts Institute of Technology, Cambridge, MA, Jun. 2010.
	
	\bibitem{CoverT1991}
	T.~M. Cover and J.~A. Thomas, \emph{Elements of Information Theory}.\hskip 1em
	plus 0.5em minus 0.4em\relax New York: John Wiley \& Sons, 1991.
	
	\bibitem{Smith1971}
	J.~G. Smith, ``The information capacity of amplitude- and variance-constrained
	scalar {G}aussian channels,'' \emph{Inf. Control}, vol.~18, no.~3, pp.
	203--219, Apr. 1971.
	
	\bibitem{Unser2000}
	M.~Unser, ``Sampling---50 years after {S}hannon,'' \emph{Proc. {IEEE}},
	vol.~88, no.~4, pp. 569--587, Apr. 2000.
	
	\bibitem{VetterliKG2014}
	M.~Vetterli, J.~Kova\v{c}evi\'{c}, and V.~K. Goyal, \emph{Foundations of Signal
		Processing}.\hskip 1em plus 0.5em minus 0.4em\relax Cambridge: Cambridge
	University Press, 2014.
	
	\bibitem{DeVoreL1993}
	R.~A. DeVore and G.~G. Lorentz, \emph{Constructive Approximation}.\hskip 1em
	plus 0.5em minus 0.4em\relax Berlin: Springer--Verlag, 1993.
	
	\bibitem{Unser1999}
	M.~Unser, ``Splines: A perfect fit for signal and image processing,''
	\emph{{IEEE} Signal Process. Mag.}, vol.~16, no.~6, pp. 22--38, 1999.
	
	\bibitem{Boor1978}
	C.~de~Boor, \emph{A Practical Guide to Splines}.\hskip 1em plus 0.5em minus
	0.4em\relax New York: Springer-Verlag, 1978.
	
	\bibitem{Tchamkerten2004}
	A.~Tchamkerten, ``On the discreteness of capacity-achieving distributions,''
	\emph{{IEEE} Trans. Inf. Theory}, vol.~50, no.~11, pp. 2773--2778, Nov. 2004.
	
	\bibitem{ElMoslimanyD2018}
	A.~El{M}oslimany and T.~M. Duman, ``On the discreteness of capacity-achieving
	distributions for fading and signal-dependent noise channels with
	amplitude-limited inputs,'' \emph{{IEEE} Trans. Inf. Theory}, vol.~64, no.~2,
	pp. 1163--1177, Feb. 2018.
	
	\bibitem{DytsoGPS2018}
	A.~Dytso, M.~Goldenbaum, H.~V. Poor, and S.~S. Shitz, ``When are discrete
	channel inputs optimal? - optimization techniques and some new results,'' in
	\emph{Proc. 52th Annu. Conf. Inf. Sci. Syst. (CISS 2018)}, Mar. 2018, pp.
	1--6.
	
	\bibitem{Kudryavtsev1995}
	S.~N. Kudryavtsev, ``Recovering a function with its derivatives from function
	values at a given number of points,'' \emph{Russian Academy of Sciences
		Izvestiya Mathematics}, vol.~45, no.~3, pp. 505--528, 1995.
	
	\bibitem{GyorfiKKW2002}
	L.~Gy{\"o}rfi, M.~Kohler, A.~Krzy{\.z}ak, and H.~Walk, \emph{A
		Distribution-Free Theory of Nonparametric Regression}.\hskip 1em plus 0.5em
	minus 0.4em\relax New York: Springer-Verlag, 2002.
	
	\bibitem{WuTVM2018_arXiv}
	T.-Y. Wu, A.~Tandon, L.~R. Varshney, and M.~Motani, ``Multicasting energy and
	information simultaneously,'' arXiv:1806.11271v1 [cs.IT]., Jun. 2018.
	
	\bibitem{Gastpar2003}
	M.~Gastpar, ``To code or not to code,'' Ph.D. dissertation, {\'{E}cole}
	Polytechnique {F\'{e}d\'{e}rale} de Lausanne, Switzerland, Jan. 2003.
	
	\bibitem{Shannon1959}
	C.~E. Shannon, ``Coding theorems for a discrete source with a fidelity
	criterion,'' in \emph{IRE Nat. Conv. Rec., Part 4}, Mar. 1959, pp. 142--163.
	
	\bibitem{ToraichiKSM1987}
	K.~Toraichi, K.~Katagishi, I.~Sekita, and R.~Mori, ``Computational complexity
	of spline interpolation,'' \emph{Int. J. Systems Sci.}, vol.~18, no.~5, pp.
	945--954, 1987.
	
\end{thebibliography}
\end{document}